\documentclass[12pt,a4paper]{article}
\usepackage{graphicx}
\usepackage{times}
\usepackage{amsmath}
\title{ \bf Seismic probe of transport processes in red giants }

\author{MarieJo Goupil$^1$\thanks{mariejo.goupil@obspm.fr}\\
  $^1$ {\small LESIA, Observatoire de Paris, 92195  Meudon, France and} \\
 {\small CNRS, PSL Research, Pierre et Marie Curie and Denis Diderot universities}}
 
\date{\mbox{}}

\def\beginrefer{\section*{References}%
\begin{quotation}\mbox{}\par}
\def\refer#1\par{{\setlength{\parindent}{-\leftmargin}\indent#1\par}}
\def\endrefer{\end{quotation}}

\begin{document}

\maketitle
\setcounter{page}{1001}
\pagestyle{plain}
    \makeatletter
    \renewcommand*{\pagenumbering}[1]{%
       \gdef\thepage{\csname @#1\endcsname\c@page}%
    }
    \makeatother
\pagenumbering{arabic}

%
%
\def\bull{\vrule height .9ex width .8ex depth -.1ex}
\makeatletter
\def\ps@plain{\let\@mkboth\gobbletwo
\def\@oddhead{}\def\@oddfoot{\hfil\scriptsize\bull\quad
"How Much do we Trust Stellar Models?", held in Li\`ege (Belgium), 10-12 September 2018 \quad\bull}%
\def\@evenhead{}\let\@evenfoot\@oddfoot}
\makeatother
{\noindent\small{\bf Abstract:} 
Seismic data obtained  with the   space photometric CoRoT and
Kepler instruments  have led to  a  unprecendently precise characterization-  in terms of masses and ages- of a large sample of post main sequence stars 
(low mass subgiant and red giants). The high quality of the collected  seismic data  and the subsequent theoretical   work for interpreting them
 brought up a series of issues  which revealed  that our  knowledge of the internal properties of red giant stars
remains quite limited. Two such important  issues are discussed here, namely  mixing beyond the convective core of helium burning red giant stars and 
  evolution of  internal angular momentum   for post main sequence stars. This includes how they were diagnosed 
and what are the resulting  improvements in our understanding  regarding these issues (or rather  how far we are from a 
 proper understanding and realistic modelling of the structure and evolution of post main sequence stars). 
} 
\vspace{0.5cm}

\section{Introduction}

Space-based observations by CoRoT (Baglin et al., 2006, 2016) and Kepler (Borucki, 2010, 2018) revealed  that low mass subgiants and red giants
(understanding here with masses below $1.4-1.5 M_\odot$) oscillate with  nonradial modes which, similarly to the Sun, are excited by the convective motions of the envelope. 
Unlike the Sun and due to their evolved structure, they actually show a rich spectrum of mixed modes 
(e.g., Dziembowski, 1971; Scuflaire, 1974; Aizenman et al., 1977). These  modes 
have significant amplitudes  both in the core - mostly
localized at and below the hydrogen-burning shell-where the dominant restoring force is  buoyancy (g-cavity)
as well as in the outer envelope where the dominant restoring force is the pressure gradient  (p-cavity)
(Dziembowski et al., 2001; Dupret et al., 2009).  This property  enables us to study  the inner properties of the star quasi directly.  
However the richness of the mixed mode frequency spectrum comes with a high complexity
  which makes  disentangling   the various modes and measurements of their frequencies highly challenging. Fortunately,
the excited range of modes  lies in a regime where the modes  closely  follow an asymptotic behavior in the g-cavity. This is less true in the 
p-cavity but deviations from asymptotics remain acceptable for first order  investigations. 
The result of such seismic analyses was tremensdouly  wealthy:  -determination of seismic masses and ages with unprecedent precision;
 -the unvaluable segregation between giant stars ascending the giant branch, stars in the primary and secundary clumps 
 which was not possible previoulsy  with classical observations only   
- measurement of the core  rotation profile  and its time evolution. These advances led to a total renewal of  the galactic archeology field.
 They also brought up a serious problem  with  the modelling of internal angular momentum transport. This  wealthy  crop of information was already reviewed (Chaplin \& Miglio, 2008; 
Mosser \& Miglio, 2016; Hekker \& Christensen-Dalsgard, 2017). Here I will concentrate  on two unsolved 
issues in the modelling of low mass subgiant and red giant stars: mixing beyond the convective core  (CC) of red clump (RC) stars 
 (Sect.3) and internal angular momentum (AM) in post main sequence stars (PoMS) (Sect.4). The first problem was already known but was confirmed 
by the  seismology of these stars while the second one was an unexpected outcome of the PoMS seismic analyses.  
Before addressing these issues, I must introduce the tools that led to these discoveries. Hence I start in Sect.2, by briefly reviewing the main seismic observables and diagnotics 
accessible to PoMS stars and the  theoretical basis that allows the interpretation of these observations. Some conclusions are drawn in Sect.5.

\begin{figure}[t]
 \centering
 \includegraphics[width=0.50\textwidth,clip]{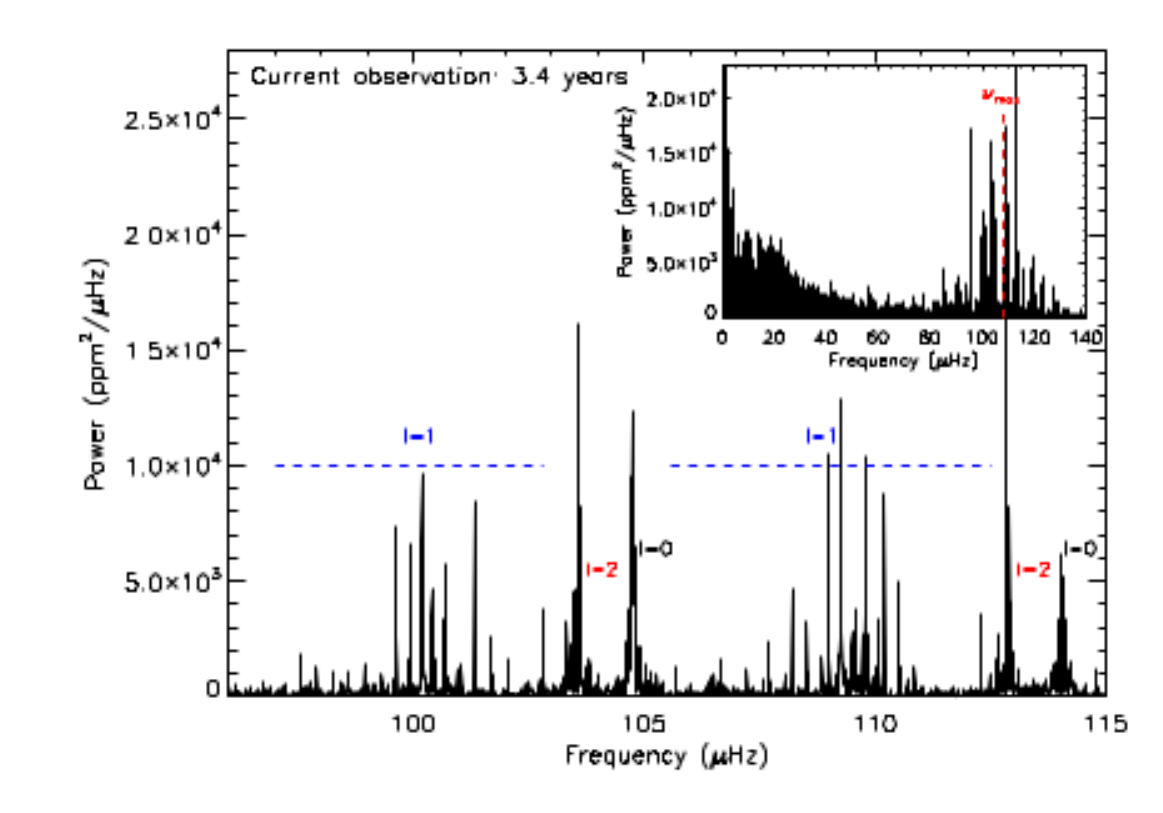}       
        \caption{\label{fig1} \small \it 
Power spectrum of the red giant KIC 10977979 from a 3.4 year-long lightcurve.   The frequency peaks associated with
 $\ell=0,1,2$ modes can be clearly seen as indicated in the figure 
  (credit Benomar et al., 2013).
  }
 \end{figure}
 
\section{Seismic  constraints for evolved low mass stars}
As already mentioned, the power spectrum of a low mass PoMS star  exhibits a complex frequency pattern.  
A typical exemple of an observed pattern of a set of $\ell=0, 1, 2$ modes is shown in Fig.\ref{fig1}.
 For sake of notation,  lets recall a few pre-requisite  general statements.  
One interprets such a frequency or power  spectrum  in a linear adiabatic normal mode framework (Cox, 1980;  Unno et al. 1989; Gough, 1993).
 The observed frequencies  are associated with normal modes of the stellar resonant cavity.   
The low mass PoMS stars are mainly slow rotators; one then usually assumes that the equilibrium state keeps its spherical symmetry.
 Accordingly the geometry of each 
 linear (eigen-)mode  can be described by means of a single spherical harmonics and the associated frequency $\nu$ and pulsation $\omega=2\pi \nu$
are usually labelled with the degree $\ell$ and the azimuthal order $m$ of the spherical harmonics. 
  In the following only $\ell=1$ modes will be considered. The $\ell, m$ indices will be omitted unless necessary.  
The frequency must also be labelled with a radial order $n$
associated with the number of nodes of the eigenfunction in the radial direction. When necessary,  one also usually
distinguishes the number of nodes in the g-cavity and  in the p-cavity and denotes them as $n_g$ and $n_p$ respectively. 
For the observed modes of evolved stars,    
$k_r ~\Delta r_{g} >>1$  in the g-cavity where $k_r$ is the radial wavenumber of the mode and $\Delta r_{g}$ the extension of the inner g-cavity 
then $n_g >>1$. In the p-cavity,  $k_r~ \Delta r_{p}  \geq 1$  with $\Delta r_{p}$ the extension of the outer p-cavity  and typically  $n_p \sim 6-10$.

A first information comes from the {\bf large frequency separation} defined  for axisymetric dipolar modes as:
\begin{equation}\label{Deltanu0}
\Delta \nu_n =\nu_n-\nu_{n-1}
\end{equation}
$\Delta \nu_n$ increases with the frequency $\nu_n$ and reaches at high frequency a nearly constant  value $\Delta \nu$ which is closely related to the 
asymptotic (theoretical)  large frequency separation  defined as 
\begin{equation}\label{Deltanu}
\Delta \nu_a =  \Bigl(2 \int_{0}^{R}~\frac{dr}{c_s}\Bigr)^{-1}
\end{equation}
where $c_s$ is the sound speed and $R$ the stellar radius.

\medskip
 A second seismic diagnostic comes from the observed {\bf period separation}, $P_n$ defined as:
\begin{equation}
\Delta P_n = P_n-P_{n-1}=\frac{1}{\nu_{n}}-\frac{1}{\nu_{n-1}}
\end{equation}
  where $P_n $  is the oscillation period of a dipole mode with radial order $n$. The period separation can be related to the asymptotic (theoretical) period spacing 
 which is defined as 
\begin{equation}\label{dPi1}
\Delta \Pi_1  = \frac{2\pi^2}{\sqrt{2}} ~ \Bigl(\int_{r_a}^{r_b} ~\frac{N}{r}~dr  \Bigr)^{-1} 
\end{equation}

The period separation is then  known to be strongly related to the evolution stage and age of the red giants (Bedding et al. 2011; Mosser et al. 2011, 2012a, 2014). 
  
\medskip
   The third main observable is the  {\bf rotational splitting}: 
\begin{equation}\label{split}
\delta \nu_n = \nu_{n,1}-\nu_{n,-1}
\end{equation} 
where the second indice refers to the azimutal number of the associated $\ell=1$  spherical harmonics. 
The measurement of  rotational splittings enabled determining the mean core rotation of subgiant and red giant stars 
(Beck et al. 2012; Mosser et al 2012b;  Deheuvels et al, 2012, 2014, 2015; Di Mauro 2016, 2018).  

\begin{figure}[t]
\begin{minipage}{8.5cm}
\centering
\includegraphics[width=8.5cm]{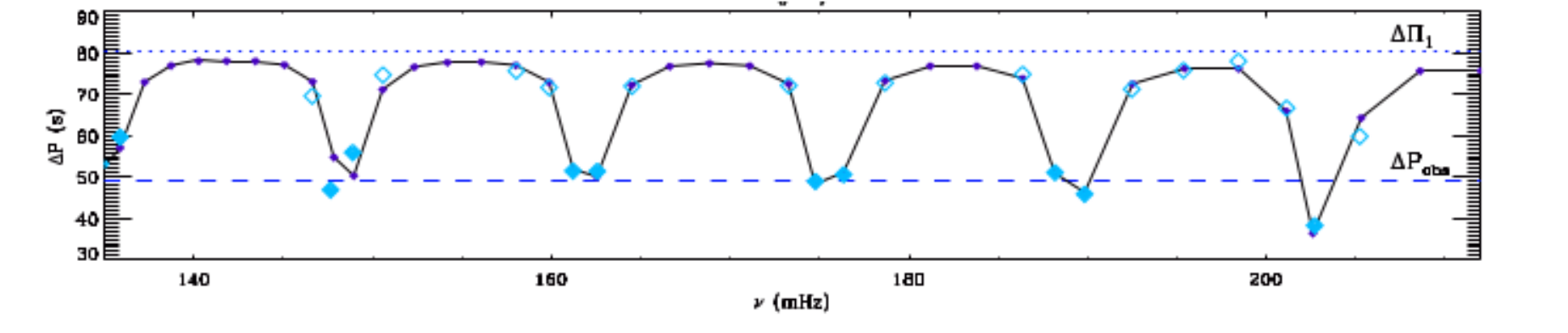}
\caption{ \it \small Period separations, $\Delta P_n$,    as a function of the frequency $\nu$ for the red giant  KIC 9882316. 
The modes with the largest period separations $\Delta P_n$ are g-dominated modes and  the modes with the smallest $\Delta P_n$  
are p-dominated  modes. 
Each cyan diamond corresponds to an observed mode of the star. The g-mode spacing  $\Delta \Pi_1$ is indicated
by the dotted horizontal line (credit Mosser et al. 2012). \label{fig2}}
\end{minipage}
\hfill
\begin{minipage}{7cm}
\centering
\includegraphics[width=7cm]{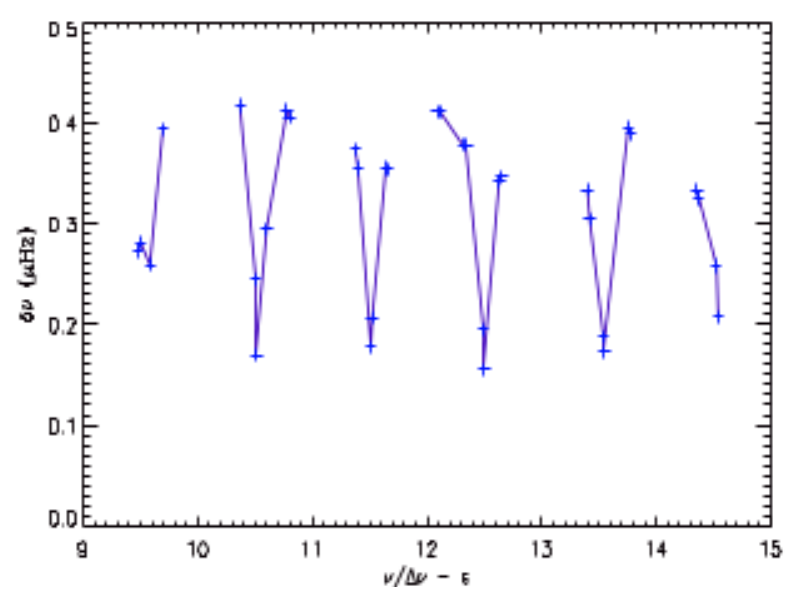}
\caption{ \it \small Rotational splittings as a function of the frequency normalized  to the mean large separation  for
the Kepler star KIC 5356201 (data from Beck et al. 2012)   (credit Goupil et al. 2013).  \label{fig3}}
\end{minipage}
\end{figure}

\medskip
Examples for these last two observables 
are displayed in Fig.~\ref{fig2} and Fig.~\ref{fig3}.

\begin{figure}[t]
 \centering
 \includegraphics[width=8cm,clip]{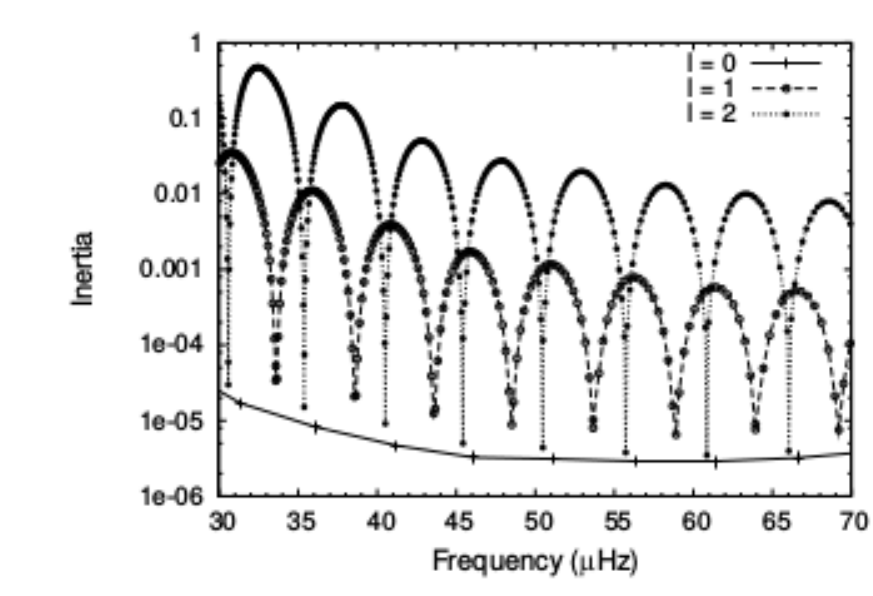}       
   \caption{\label{fig4} \small \it 
  Mode inertia  for a red giant model as a function of the mode frequency for $l=0, 1,2 $. Highest inertia are g-dominated modes. 
Lowest inertia are p-dominated modes. Intermediate inertia correspond to intermediate cases. 
 The lowest curve connecting the dots represent the inertia of radial modes (credit Dupret et al., 2009). 
}
 \end{figure}

\medskip 
Before presenting the fourth observable, one must introduce  mode inertia, $I(\nu)$, which is defined as:
 \begin{equation}
I(\nu) =   \int_0^R \Bigl(\xi_r^2+ \Lambda \xi_h^2\Bigr) ~  4\pi\rho r^2 dr
 \end{equation}
where $\Lambda = \ell(\ell+1)$  and $\xi_r$ and  $\xi_h$ represent the  vertical and horizontal displacement eigenfunctions 
respectively (Dziembowski et al., 2001).  Fig.~\ref{fig4} shows that mode inertia  significantly changes
 for mixed modes lying between two successives radial modes.  Mode inertia, $I$,   is  much 
larger for g-dominated modes than p-dominated modes due to the  inward increase of the  mass distribution.

\begin{figure}[t]
 \centering
 \includegraphics[width=12cm,clip]{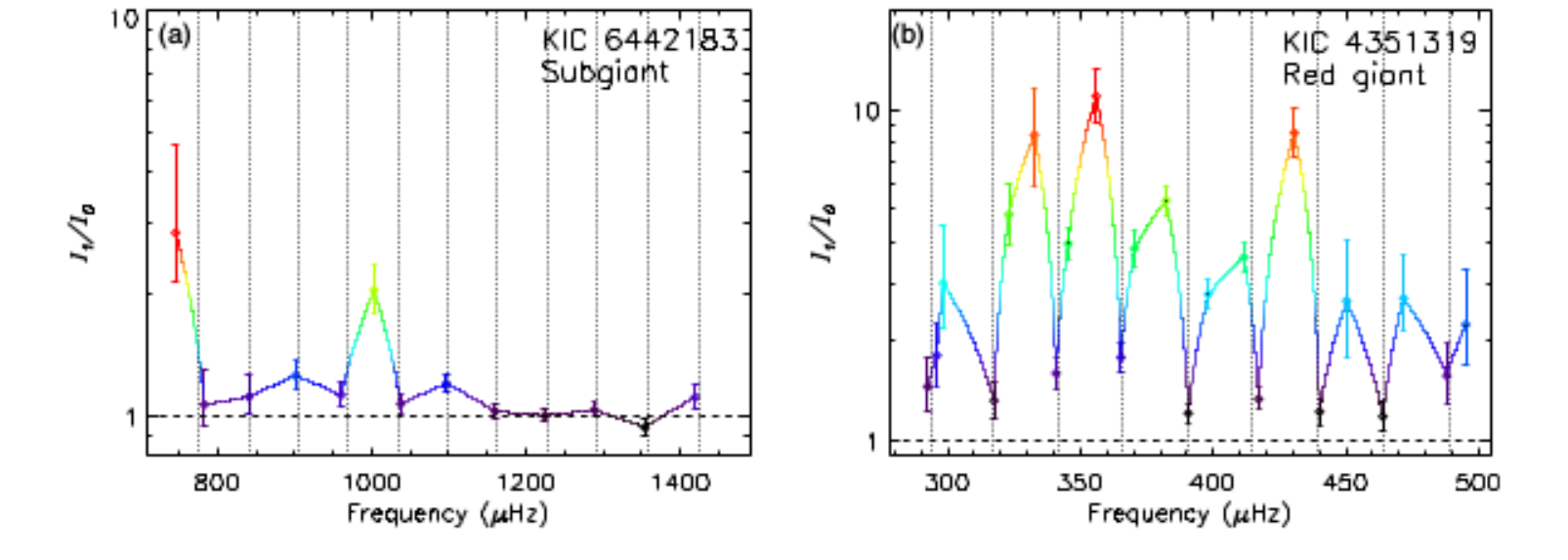}       
   \caption{\label{fig5} \small \it 
  Observed core-to-total inertia ratio  for the red giants (a) KIC 6442183 and (b) KIC 4351319 against frequency. 
The black, blue, red, and brown diamonds are symbols for $\ell=0, 1,2,3$ respectively.
 Vertical dashed lines indicate the position of the nearly pure p-modes 
(from Benomar et al. 2014).  }
 \end{figure}

\medskip
The fourth observable discussed hereafter then is   the {\bf inertia ratio} (Goupil et al. 2013; Deheuvels et al. 2015; Benomar et al., 2014) defined as: 
\begin{equation}\label{zeta}
\zeta (\nu)= \frac{I_{core}}{I}
\end{equation}
The observed variations of $\zeta$ with frequency are displayed in Fig.\ref{fig5} for two red giants.
  A good approximation is   $I \approx I_{core}+I_{env}$ where one identifies the core with the g-mode cavity and 
the envelope with the p-mode cavity (neglecting inertia in the intermediate evanescent region). 
Accordingly,  $\zeta \approx 1$ for g-dominated modes and  $\zeta <<1$ for p-dominated modes. 
  
Interpretation of the observed period spacings and rotational splittings are 
easier when one realizes that these observables are both linearly  related to mode inertia. More precisely, they 
follow the same behavior as a function of frequency  than the  inertia ratio  $\zeta$ (Eq.\ref{zeta}).
 This is discussed in more detailed in the next section.

\begin{figure}[t]
 \centering
 \includegraphics[width=0.70\textwidth,clip]{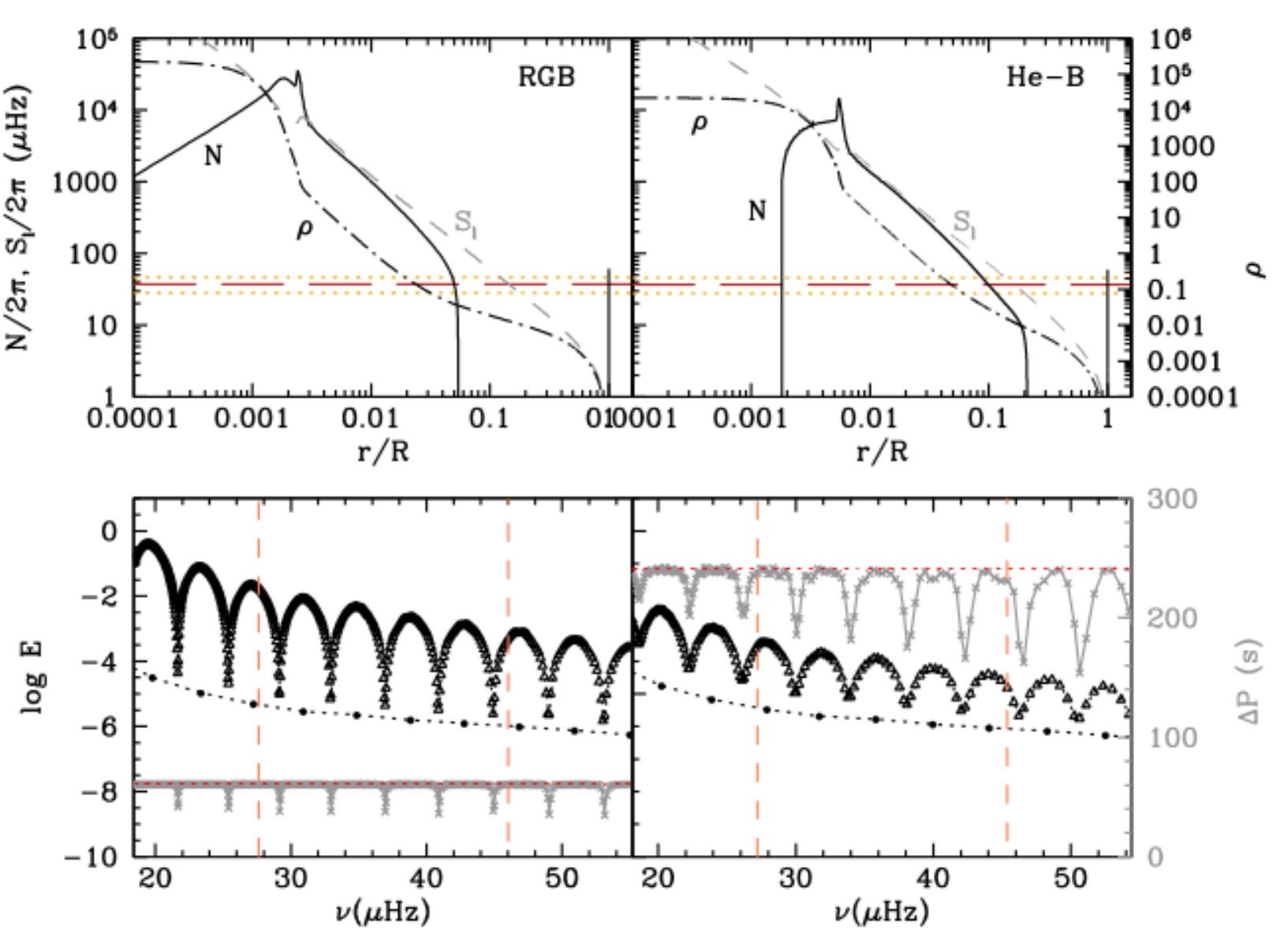}       
   \caption{\label{fig6} \it \small 
Top: Characteristics frequencies  $N,S_1$ and density distribution for dipole modes for  
 a  $1.5 M_\odot, 12 R_\odot$ stellar model for a red giant (left) and a red clump star (right).
The typical  frequency range of excited modes for each star is  shown by the horizontal dotted lines.
 The three turning points  ($k_r=0$) for these models 
are located at the intersection of this range and either $N$ or $S_1$.
 Bottom: inertia as a function of the frequency for
$\ell=0$ (circles) and 1 (triangles) modes. The vertical dashed  lines represent the typical  frequency range of excited modes for such stars.
 Highest inertia are g-dominated modes. Lowest inertia are p-dominated modes.
Grey crosses connected with a dotted line represent the period separation $\Delta P_n=P_{n+1}-P_{n}$. 
The horizontal red thin-dashed line indicates the asymptotic period spacing $\Delta \Pi_1$ (credit Montalban et al., 2013).
   }
 \end{figure}

\subsection {Asymptotics of mixed dipole modes  for post main sequence stars}

 The  properties of the observed oscillations of subgiants and red giants can be well investigated  with an asymptotic analysis of mixed modes. 
In that framework, the  structure of the equilibrium model is represented by two characteristic frequencies,  the Lamb, $S_\ell$, and 
the Br\"unt-V\"aiss\"al\"a, $N$, frequencies    defined respectively as 
  \begin{equation}
 S_\ell^2 = \frac{\Lambda~c_s^2 }{r^2} ~~~~~~;~~~~~ N^2  = g ~\Bigl( \frac{1}{\Gamma_1} \frac{d\ln p}{dr}-\frac{d\ln \rho}{dr} \Bigr) 
\end{equation} 
where $c_s$ is the sound speed; $\Lambda = \ell(\ell+1)$, $\Gamma_1,g, p,\rho$ have their usual meanings
  (see for instance Montalban et al., 2013 and Montalban \& Noels, 2014).
In the following, the  inner g-cavity is located between the radii $r_a$ and  $r_b$  ($r_a \sim 0 << r_b$).
 The outer p-cavity is delimited by the radii  $r_c$ and $r_d$  ($r_b< r_c << r_d \sim R$).  These resonant cavities
 are separated by an intermediate evanescent region (delimited by the radii $r_b, r_c$). 
The situation is well depicted  in Fig.\ref{fig6} for a red giant and  a subgiant  star. 

The  resonant condition for those modes was first derived by Shibahashi (1979) (see also Unno et al (1989)) in the Cowling approximation  and 
in the limit of a thick evanescent region
\begin{equation}\label{cond1}
 \cot \theta_g ~ \tan \theta_p = q
\end{equation}
where 
 \begin{equation} \label{theta}
 \theta_p =\int_{r_c}^{r_d} ~ k_r(r,\omega) ~~dr   ~~~~~~; ~~~~~ \theta_g  =\int_{r_a}^{r_b} ~ k_r(r,\omega)~~ dr 
 \end{equation}
and $q$ is the coupling  factor between the two trapping regions. 

When contributions from regions close to the turning points of the resonant cavities are neglected, 
 the radial wave number $k_r (r,\omega)$  is   given by 
  \begin{equation}
 k_r^2 (r,\omega)\approx  \frac{1}{\omega^2 c_s^2} \Bigl(\omega^2-S_\ell^2\Bigr) ~   \Bigl(\omega^2-  N^2 \Bigr) 
 \end{equation}
The turning points ($k_r=0$) $r_a,r_b,r_c,r_d$ are then given by  $\omega = S_1(r_c)= S_1(r_d) =N(r_a)=N(r_b)$.
 
\medskip
\paragraph{G-cavity: } Away from the turning points, $\omega <<S_l,N$. The phase $\theta_g$ for the g-cavity can be approximated as: 
\begin{equation}\label{thetag}
\theta_g \approx \sqrt{\Lambda} ~ \int_{r_a}^{r_b} ~  \Bigr(\frac{N^2}{\omega^2}-1\Bigl)^{1/2} ~ \frac{dr}{r} 
 \approx \frac{\pi}{\nu ~\Delta \Pi_1}  + \pi \epsilon_g(r_a,r_b,\omega)
\end{equation}
where $\nu=2\pi/\omega$.

For a pure g-mode trapped   in a g-cavity delimited by $r_a,r_b$, the resonant condition  imposes:
$$\int_{r_a}^{r_b} ~ k_r(r,\omega)~~ dr =  \frac{\pi}{\nu ~\Delta \Pi_1}  + \pi \epsilon_g(r_a,r_b,\omega) = n_g \pi$$
 
For a g-dominated  mixed mode,  the influence of the p-cavity  rather leads to define  a frequency $\nu_g$ such that:
 \begin{equation}\label{nug} 
\int_{r_a}^{r_b} ~ k_r(r,\omega_g)~~ dr =  \frac{\pi}{\nu_g ~\Delta \Pi_1}  + \pi \epsilon_g(r_a,r_b,\omega_g) \approx \Bigl(n_g +\frac{1}{2}\Bigr)\pi
\end{equation}

 For later purposes, the phase $\theta_g $, Eq.\ref{thetag},  is then conveniently rewritten as 
  \begin{equation}\label{thetag2}
 \theta_g \approx \frac{\pi}{\Delta \Pi_1} ~\Bigl(\frac{1}{\nu} - \frac{1}{\nu_g}   \Bigr) +  \Bigl(n_g +\frac{1}{2}\Bigr)\pi  
\end{equation}

\medskip
\paragraph{P-cavity:} Away from the turning points, $\omega >> S_l,N$ and
the  phase in the p-cavity is  written as :

\begin{equation}\label{thetap}
\theta_p \approx   2\pi \nu~ \int_{r_c}^{r_d} ~   \Bigr(1-\frac{S_1^2}{\omega^2}\Bigl)^{1/2}  ~ \frac{dr}{c_s} + \pi \epsilon_p(r_c,r_d,\omega_p)
\end{equation}

We now define $\nu_p$  the frequency  of the mode as  if it was not perturbed by the g-part i.e. given by the acoustic resonant condition  
  \begin{equation}\label{nup} 2\pi \nu_p~\int_{r_c}^{r_d}  \Bigr(1-\frac{S_1^2}{\omega^2} \Bigl)^{1/2}  ~ \frac{dr}{c_s}
 + \pi \epsilon_p(r_c,r_d,\omega_p)\approx  n_p \pi
\end{equation}

For later purposes, the phase $\theta_p $, Eq.\ref{thetap} is then conveniently rewritten as: 
\begin{equation}\label{thetap2}
\theta_p \approx  2\pi (\nu-\nu_p) ~ \int_{r_c}^{r_d} ~   \Bigr(1-\frac{S_1^2}{\omega^2}\Bigl)^{1/2}  ~ \frac{dr}{c_s}+ n_p \pi 
\approx \frac{\pi}{\Delta \nu} (\nu-\nu_p) + n_p \pi 
\end{equation}
 
   \medskip

The phases  $\epsilon_g(r_a,r_b,\omega)$ and   $\epsilon_p(r_c,r_d,\omega)$  
take  into account the complex behavior close to the turning points and possible deviations from asymptotics. 
The phase $\epsilon_g$ was investigated theoretically by Takata et al. (2016a,b), Pincon et al. (2019) and observationally by Mosser et al. (2018) and 
Hekker et al. (2018).
In a  simplifying approach, Shibahashi (1979) neglected  these complexities and disgarded the phases 
 $\epsilon_g(r_a,r_b,\omega)$ and   $\epsilon_p(r_c,r_d,\omega)$.  

\paragraph{Coupling factor:} 
According to Shibahashi's approach,  when the evanescent region is thick  i.e.  when
$r_c-r_b >> \Bigl(\int_{r_c}^{r_d}~   |k_r| ~dr \Bigr)^{-1} $,  the coupling factor is given to a good approximation by: 
\begin{equation}\label{coupling}
q =\frac{1}{4}~e^{-2 \theta_e} \approx \frac{1}{4}  \Bigl(\frac{r_c}{r_b} \Bigr)^{-2}   << \frac{1}{4}\end{equation}
where the phase for the evanescent region is given by:
\begin{equation}\label{thetae}
\theta_e = \int_{r_b}^{r_c} ~|k_r| ~dr
\end{equation}
In the  other limit,   the coupling factor  for a thin evanescent region is  given 
Takata (2016a,b) by:
$$q = \frac{1- R}{1+R} $$
where $R$ is  the  wave reflexion  factor.  For a thin evanescent region, $R <<1$  and $q \sim 1$.

 For sake of simplicity  in the following,   we take $q$ independent of the frequency.

\medskip
The resonant condition can take an alterative and equivalent form.  Following Jiang \& Christensen-Dalsgaard (2014) and Cunha et al (2015),
 Eq.\ref{cond1} is rewritten as 
  $$    \cos \theta_g \sin  \theta_p - q \sin\theta_g \cos \theta_p  = 0 $$       which can be cast under the form 
\begin{equation}\label{cond2} 
 C(\omega) ~\cos (\theta_g + \phi)= 0
 \end{equation}
where $C(\omega)$ and $\phi$ are defined such that 
$ C(\omega) \cos \phi= \sin \theta_p$ and $ C(\omega) \sin \phi= q~  \cos \theta_p $ and 
 \begin{equation}\label{tgphi}
\phi = \tan^{-1} \Bigl( \frac{q}{\tan \theta_p}\Bigr)
 \end{equation}
 From Eq.\ref{cond2} the resonant condition takes  perhaps a more familiar form: 
 \begin{equation}\label{cond3}
  \int_{r_a}^{r_b}  ~ k_r~ dr  = (n_g+\frac{1}{2}) \pi - \phi 
\end{equation}
i.e. this can be interpreted as the resonant condition for a g-mode perturbed by the p-mode cavity (represented by the phase $\phi-\pi/2$).

\medskip

The resonant condition  Eq.\ref{cond1}  was found to be  amazingly useful  to characterize mixed modes of post main sequence stars
 (see for instance Mosser  et al. 2012, 2015, Deheuvels et al. 2012, 2014, 2015, Goupil et al 2013, Vrard et al.  2016, Mosser et al 2018). 
For a comprehensive review, see Hekker \& Christensen-Dalsgaard (2017).

\begin{figure}[t]
 \centering
  \includegraphics[width=8cm,clip]{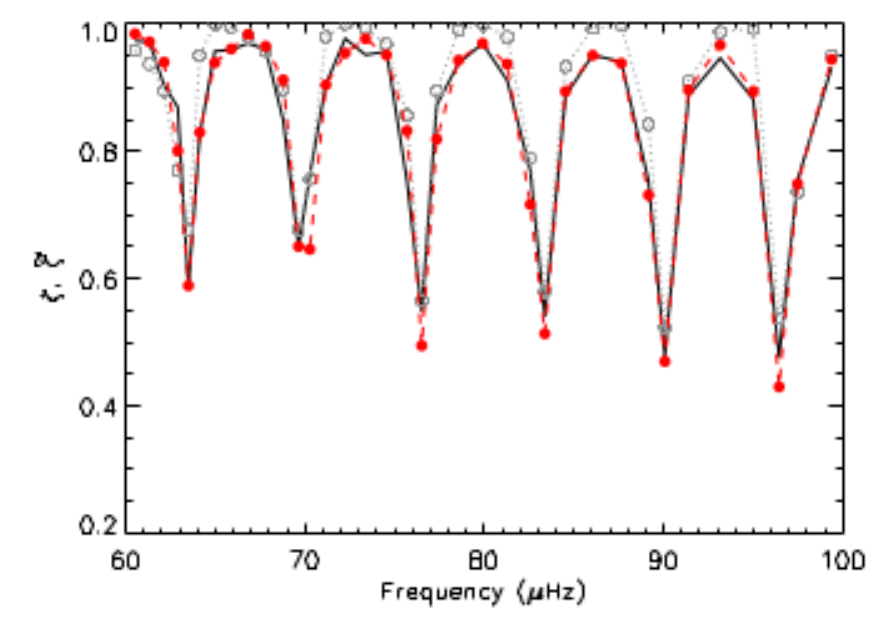} 
   \caption{\label{fig7} \small \it 
 Inertia ratio $\zeta$ for   $\ell = 1$ modes of  a model for the Kepler star  KIC7581399  computed from the numerical 
 eigenfunctions  (black solid line) and from approximate analytical expressions  with various degrees of approximation
(red filled circles and red dashed line; gray dotted line and open circles) (credit Deheuvels et al 2015).}
 \end{figure}

\paragraph{An expression for the inertia ratio, $\zeta$:}
Insights in the propagative properties  of the modes  can be obtained with an investigation of the behavior of the  inertia ratio, $\zeta$ (Eq.\ref{zeta}). 
Based on  Shibahashi (1979)'s approach, an expression for the  inertia ratio  was derived
 (Goupil et al., 2013, Deheuvels et al. 2015, Mosser et al. 2015) as:  
 \begin{equation}\label{zeta2}
\zeta (\nu)=  \Bigl(1+ \frac{1}{q} ~ \frac{\cos^2  \theta_g }{\cos^2 \theta_p }
~ \frac{\Delta \Pi_1 ~ \nu^2}{\Delta \nu} \Bigr)^{-1} 
\end{equation}
where $\theta_g$, $\theta_p$, $\Delta \Pi_1$, $\nu_p$ and $\Delta \nu$ are defined  in  Eq.\ref{thetag2}, Eq.\ref{thetap2}, Eq.\ref{nup} 
and Eq.\ref{Deltanu0} respectively. 

For a given stellar model,  $\zeta(\nu)$  (Eq.\ref{zeta}) can be computed using   numerical eigenfunctions,
 solutions of the oscillation equations (Unno et al., 1989). The comparison with the analytical approximate formulation, Eq.\ref{zeta2}, 
shows a good agreement. 
The $\zeta$ variation from one mode to the next one is represented as a function of frequency  in Fig.\ref{fig7} for  a
 model for the Kepler star  KIC7581399.   Fig.\ref{fig7} shows that the values of the local maxima of
 $\zeta$  (associated with g-dominated modes)  do not vary much  with frequency  
 whereas  the local minima (associated to p-dominated modes)  values decrease  with increasing frequencies. 

\begin{figure}[t]
\centering
\includegraphics[width=8cm]{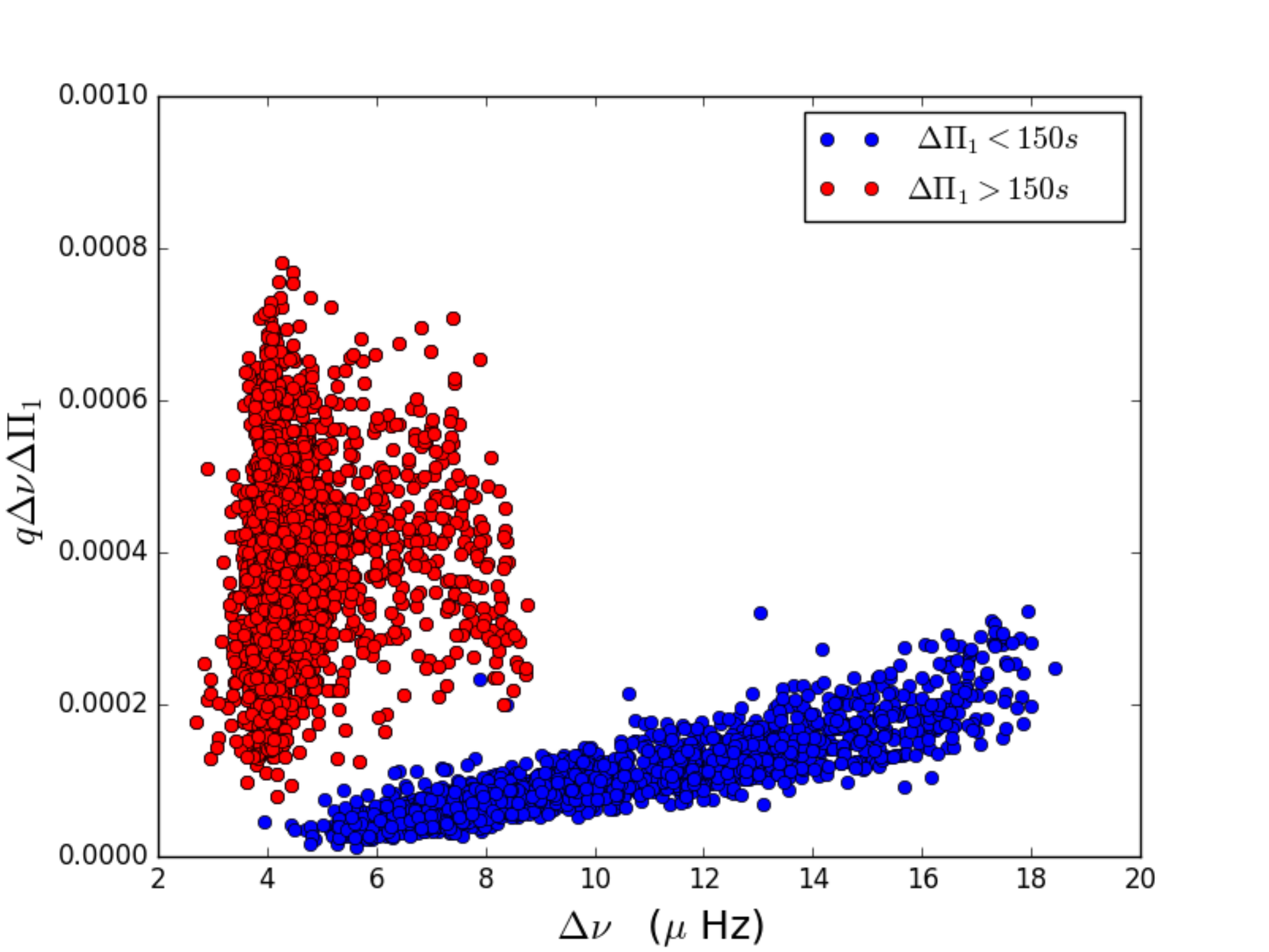}
\includegraphics[width=8.cm]{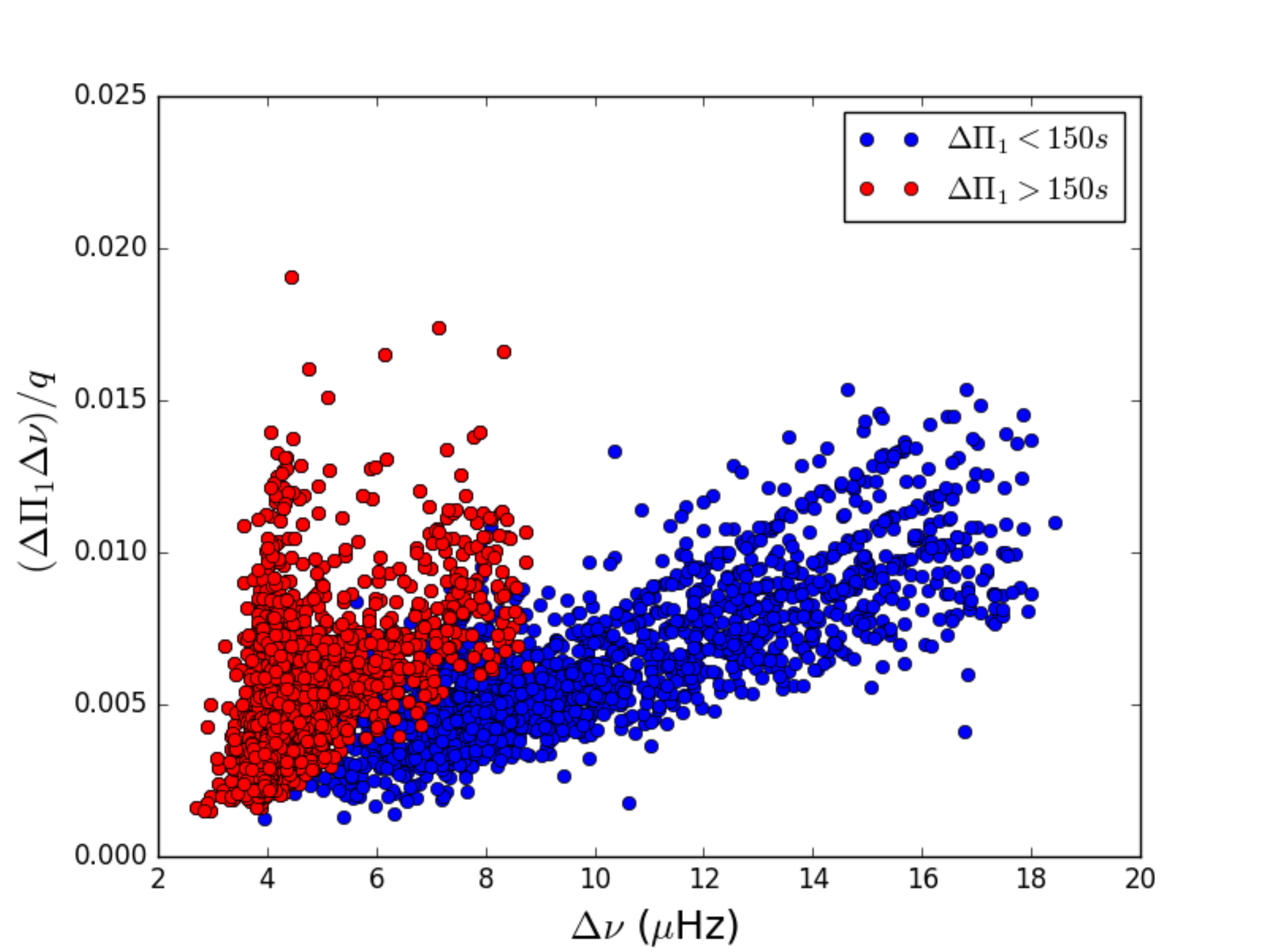}
\caption{ \it \small  \label{fig8} left: values of the factor    $q \Delta \Pi_1 \Delta \nu$ 
 as a function of $\Delta \nu$ computed for the 
 data set from Mosser et al. (2017). Stars with $\Delta \Pi_1 < 150$  roughly correspond to   RGB stars while stars with 
$\Delta \Pi_1 > 150$ s are clump stars. 
Right:  same as left for the factor    $(\Delta \Pi_1 \Delta \nu)/q$}.
\end{figure}

 The  local  maxima and  minima of  $\zeta$ are given by (Mosser et al., 2015):
\begin{equation}\label{zetaminmax}
 \zeta_{min} \approx   \Bigl(1+ \frac{1}{q {\cal N} }   \Bigr)^{-1} ~~~~~; ~~\zeta_{max} \approx   \Bigl(1+ \frac{q}{ {\cal N}}  \Bigr)^{-1}
\end{equation}
where 
\begin{equation}\label{calN}
{ \cal N}  \approx  \frac{\Delta \nu}{\Delta \Pi_1} \frac{1}{\nu^2} = \frac{1}{\Delta \nu ~\Delta \Pi_1}
 ~\Bigl(\frac{\Delta \nu}{\nu}\Bigr)^2
 \end{equation}
The coupling factor, $q$, takes  values in the range (0,1).   
The non dimensional quantity $1/(\Delta \nu ~\Delta \Pi_1) $  corresponds to the number of mixed modes 
 sitting between two consecutive p-dominated modes within a $\Delta \nu $ frequency interval.  This quantity usually is much larger than 1. 
From Eq.\ref{calN}, we have: 
\begin{equation}
\frac{q}{ {\cal N}}  \approx  q  \Delta \nu \Delta \Pi_1 n_p^2 ~~~;~~~ \frac{1}{q {\cal N}}  \approx \frac{1}{q   }  \Delta \nu \Delta \Pi_1 n_p^2
 \end{equation}
where $n_p =\nu/\Delta \nu$. The factors $q  \Delta \nu \Delta \Pi_1$ and $ \Delta \nu \Delta \Pi_1/ q$   are displayed in Fig.\ref{fig8} 
as a function of $\Delta \nu$  for the Mosser et al 2017's  sample of stars.
 The factor $q  \Delta \nu \Delta \Pi_1$ remains smaller than $4.10^{-3}$ for the RGB star sample of interest here. 
For RGB stars, these factors  tend to increase  with evolution represented by $\Delta \nu$ here.
 Further, for the detected modes in the vicinity of $\nu_{max}$
 corresponding to the maximum of the  power spectrum, 
$ n_p \sim 6-10 $ for the sample of RGB stars studied by Mosser et al. (2017), Vrard et al. (2016). 
We then expect
 ${\cal N} >>1$  and $q/{\cal N} <<1 $ which implies  that $\zeta_{max}\sim 1$.

From Eq.\ref{zetaminmax}, one derives the $\zeta$ ratio as:
\begin{equation}\label{zetaminzetamax2}
 \frac{\zeta_{min}}{\zeta_{max}} \approx \Bigl(1+ \frac{1}{q {\cal N} }\Bigr)^{-1} ~\Bigl(1+ \frac{q}{ {\cal N}}  \Bigr)
\sim  \frac{1}{1+ \frac{ \displaystyle \Delta \nu\Delta \Pi_1}{\displaystyle q } n_p^2} \end{equation}
For RGB stars, this ratio tends to decrease with $n_p^2$  with a rate
  given by the ratio $ q/(\Delta \nu\Delta \Pi_1)$.

As an illustration, I consider the  RGB star, KIC 4448777, studied by M.P Di Mauro and collaborators (Di Mauro  et al, 2018).
Two best fitting models give the  mass and radius  of the star
 to be either (1.0,3.94) or (1.13, 4.08)  in solar units,  respectively  (Di Mauro  et al, 2018) . 
Mosser et al. (2018) and Vrard et al. (2018) 
provided the values of the    characteristic quantities for that star that-is
$q=0.14\pm 0.05, \Delta \nu= 16.95-17.01~\mu Hz, \Delta \Pi_1=89.3\pm 1.76$s. The radial order 
 in the vicinity of  $\nu_{max}$ is $n_p = 11$.  
  With these values, one obtains:
$q/ {\cal N}= q ~\Delta \nu \Delta \Pi_1~ n_p^2 =2.12~10^{-4} ~ n_p^2 = 7.65~10^{-3}- 4.78~10^{-2} <<1$ 
 for $n_p$ in the range (6,15) and  $1/(q {\cal N})  \sim    0.011 n_p^2 \sim 0.39-2.44$ and
$\zeta_{min}/\zeta_{max} \sim  0.44 -0.0022 (n_p^2 -121)$.

\begin{figure}[t]
\begin{minipage}{8.cm}
 \centering
\includegraphics[width=7cm,clip]{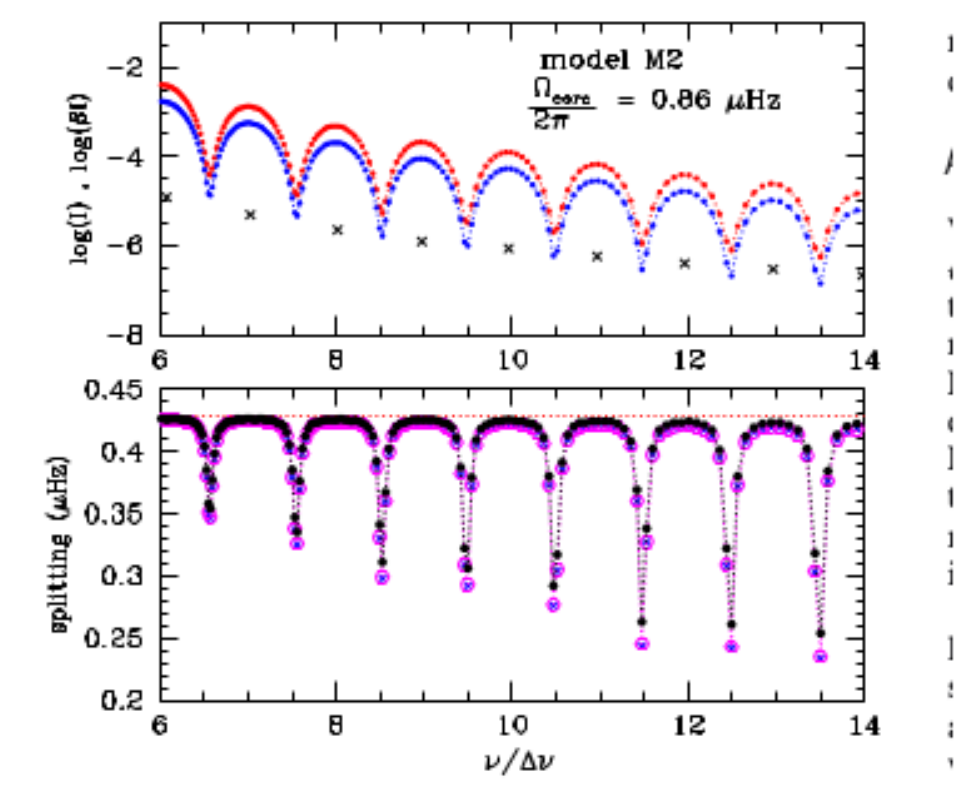}       
   \caption{\label{fig9}  \small  \it Top: inertia ratio  as a function of frequency numerically computed for a 1.3 $M_\odot$ red giant model. 
 Crosses, red and blue dots represent inertia ratios for $\ell=0,1,2$ respectively. Bottom: 
theoretical rotational splitting as a function the normalized frequencies of  (credit Goupil et al, 2013). }
 \end{minipage}
\hfill
\begin{minipage}{8.cm}
\centering
\includegraphics[width=8cm,clip]{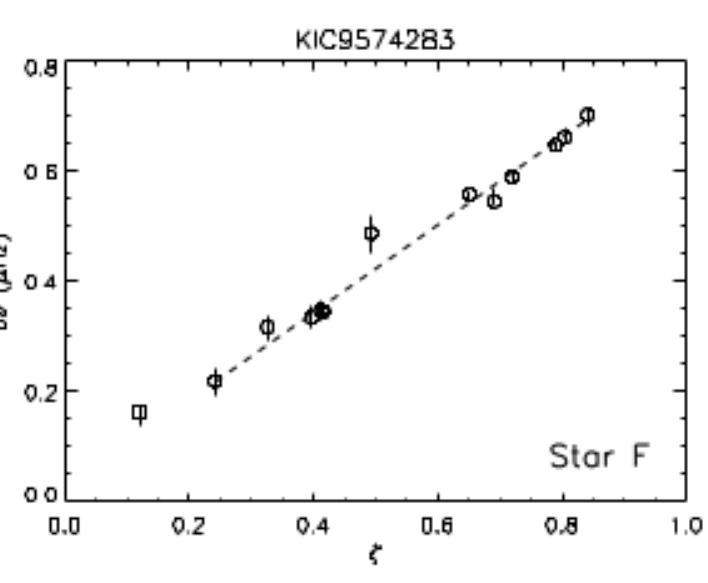}       
\caption{\label{fig10} \small \it Observed rotational splittings as a function of the  inertia ratio for the Kepler star KIC8751420. 
(circles for $\ell=1$ modes and squares for $\ell=2$ modes). The black dot-dashed line shows a linear regression of the observed 
splittings as a function of $\zeta$  
  (credit Deheuvels et al., 2014).}\end{minipage}
 \end{figure}

\paragraph {Rotational splitting:}

The theoretical rotational splitting is defined as 
\begin{equation}\label{split}
\delta \nu_{n,m} = m~ \int_{0}^{R}~  K_n(r)~ \frac{\Omega(r)}{2\pi}~ dr
 \end{equation}
where the angular dependence has been assumed averaged out to keep only the radial dependence. The kernel 
$K_n(r)$ involves the eigenfunction and structure of the star through the pressure and other profiles (Christensen-Dalsgaard, 2007). 

Fig.\ref{fig9} shows the rotational splittings computed from Eq.\ref{split} with the numerically computed eigenfunctions and 
assuming a rotation profile decreasing  outward. The  local maxima  of the splittings are those associated with g-dominated modes while the local minima
 correspond  to the p-dominated modes. 
Measurements of the splitting maximum provides the core rotation rate (Beck et al. 2012, Mosser et al. 2012b).
\begin{equation}\label{numax1}
\delta \nu_{max} \approx \frac{ \Omega_{core}}{4\pi}
\end{equation}
 
\begin{figure}[t]
 \centering
\includegraphics[width=8cm,clip]{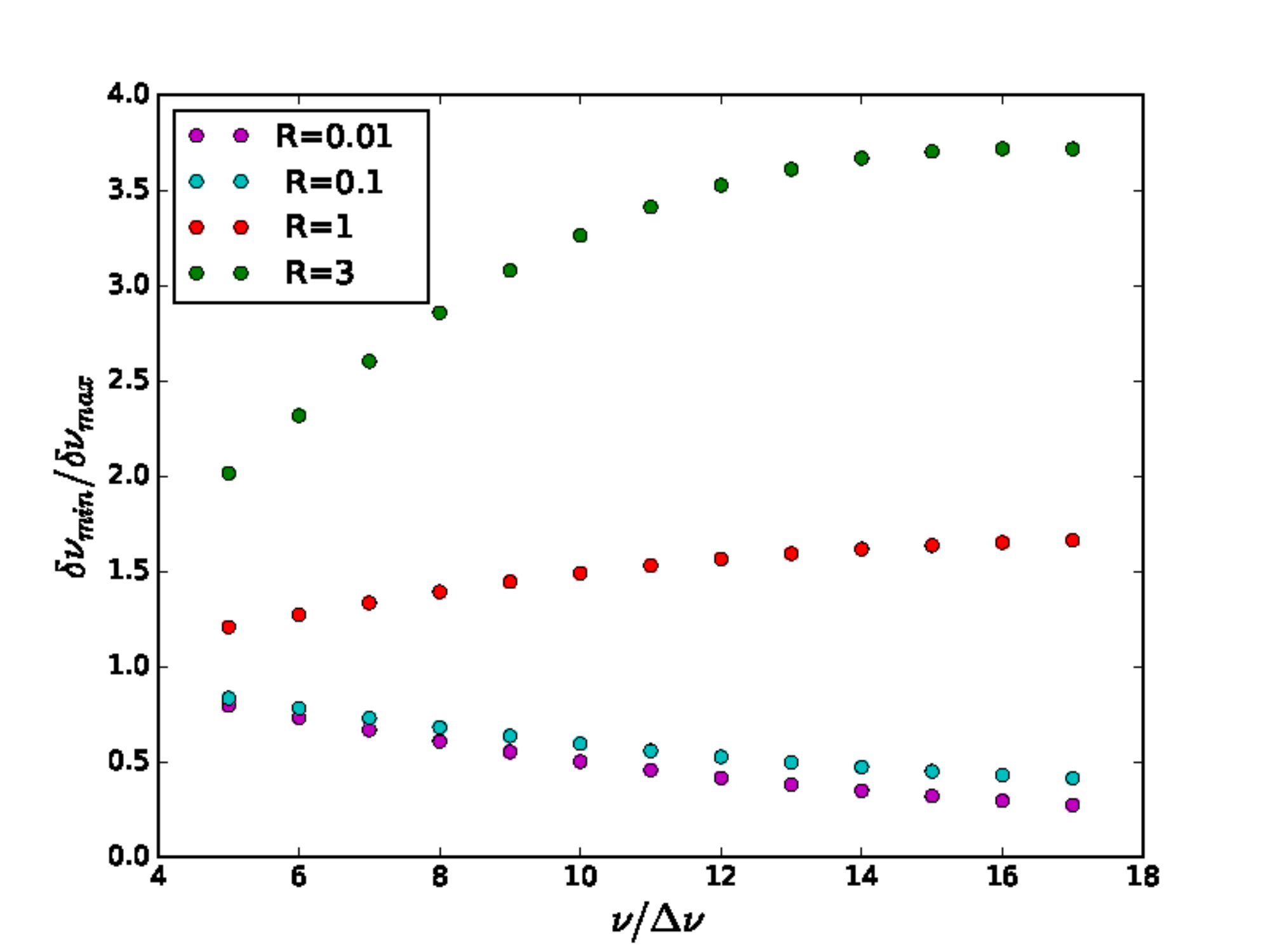}       
 \includegraphics[width=8cm,clip]{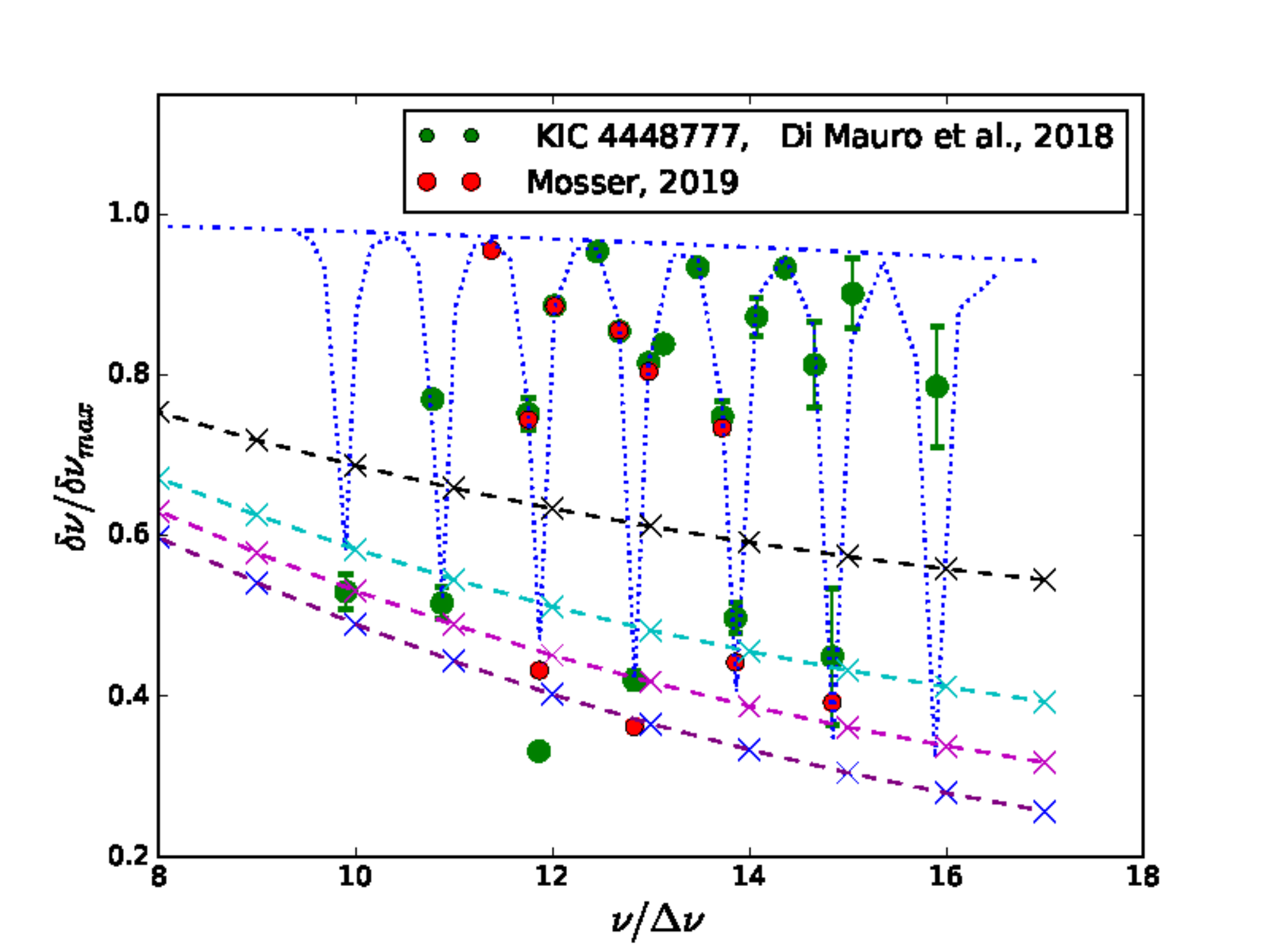}       
   \caption{\label{KIC444} \small \it  left:
Splitting ratio as a function of the radial order $\nu/\Delta \nu$ for several values of the rotation ratio ${\cal R}$. 
Right: observed splittings for KIC 448777  normalized to the constant value $0.4 ~\mu Hz$   as a function of 
 $\nu/\Delta \nu$ . The  dotted curve represents $\delta \nu_{max}/0.4 = 1/(1+q {\cal N}) $.  
  Crosses connected with dashed lines  show the discrete values of the ratios 
$(1+2 {\cal R}/(q {\cal N}) / (1+1/(q {\cal N})) $ for 
$  {\cal R}=0.2$ (black), $ {\cal R}=0.1 $ (cyan),  $ {\cal R}=0.05 $ (magenta),  $ {\cal R}=0.01$ (purple).}
 \end{figure}

Further  Goupil et al. (2013)   showed that  the rotational splittings   follow the behavior of  the inertia ratio $\zeta$ (Fig.\ref{fig10}).
From Eq.15, Eq.A3 and Eq.A6 of that Goupil et al. (2013)'s paper,  the  splitting can be written    as: 
\begin{equation}\label{deltanu}
 \delta \nu \approx (1-\zeta) \frac{\Omega_{env}}{2\pi} + \frac{\zeta}{2} \frac{\Omega_{core}}{2\pi}
 \end{equation}
Then the local maxima are 
\begin{equation}
\label{numax2}
 \delta \nu_{max} \approx  \frac{\zeta_{max}}{2} \frac{\Omega_{core}}{2\pi} \Bigl(1+ 2\frac{(1-\zeta_{max})}{\zeta_{max}} {\cal R}  \Bigr) 
\approx \frac{1}{2} \frac{\Omega_{core}}{2\pi}  \frac{1+2 (q/{\cal N}) {\cal R}}{1+ q/{\cal N} }
\end{equation}
where Eq.\ref{zetaminmax}  was used  and $\Omega_{env}$ and $\Omega_{core}$  represent 
seimic averages of the rotation rate over the core and over the envelope respectively; 
${\cal R} = \Omega_{env}/\Omega_{core}$ is the rotation ratio or 'gradient'. 
 Stellar models predict that ${\cal R}$ decreases when the star ascends the RGB. 
 For the most g-dominated modes, $q/{\cal N}<<1$ then  one recovers Eq.\ref{numax1}. Actually, $\delta \nu_{max}$ slightly decreases with frequency. 
 Eq.\ref{numax2} well 
reproduces this decrease of the local maxima with the frequency. An example is shown in Fig.\ref{KIC444}. 

Since individual $ \delta \nu $ and $\zeta(\nu)$ are observable, one has access to  a precise measurement of 
 $\Omega_{core}$ (Eq.\ref{numax1} or Eq.\ref{numax2}) and -provided  that  p dominated modes are detectable-
  to  $\Omega_{env}$ or  $\Omega_{env}/\Omega_{core}$ (Eq.\ref{deltanu}). 
This was validated by comparison with results of  numerical models  and observations.

 As shown by Klion et al. (2017), the ratio  ${\cal R} $
depends on  the location of the differential rotation within the star.  ${\cal R} $ can be obtained at a given frequency. 
 Klion et al. (2017) suggested  to use $\delta \nu_{min}/\delta \nu_{max} $ in the vicinity of 
 $\nu_{max}$, the frequency at maximum power, in order  to obtain information about the location of the differential rotation. 
Setting $n_p$ to $n_{p,max}= \nu_{max}/\Delta \nu$, the splitting ratio then only depends on 
 ${\cal R}=\Omega_{env}/\Omega_{core}$ and on the observable  $  \Delta \nu \Delta \Pi_1/q $ for each star. 
A plot of $\delta \nu_{min}/\delta \nu_{max}$ as a function of  
$  \Delta \nu \Delta \Pi_1/q$ provides the rotation gradient ${\cal R} $.  
Following this idea, a general fit can also in theory provides the ratio ${\cal R}$. 
Using Eq.\ref{deltanu} and Eq.\ref{zetaminmax}, one derives the following expression for the splitting ratio:
\begin{equation}\label{dnumindnumax}
\frac{\delta \nu_{min}}{\delta \nu_{max}} \approx
  \frac{ (1-\zeta_{min})  \Omega_{env}  + (\zeta_{min}/2)~  \Omega_{core} }
{(1-\zeta_{max})  \Omega_{env}  + (\zeta_{max}/2) \Omega_{core} } =    
 \Bigl(\frac{1+  \frac{\displaystyle q}{\displaystyle{\cal N}}     } {1+     \frac{\displaystyle  1}{\displaystyle q {\cal N} } }  \Bigr)  ~
\Bigl( \frac{1+ 2    \frac{\displaystyle  1 }{\displaystyle q {\cal N} }\cal R }{1+2~   \frac{\displaystyle q}{{\displaystyle \cal N}}    {\cal R}  } \Bigr)
\end{equation}
  
In Fig.\ref{KIC444}(left), the splitting ratio are plotted as a function of the radial order $n_p =\nu/\Delta \nu$ 
for several values of the rotation ratio ${\cal R}$. For large differential rotation ${\cal R}$, the splitting ratio increases with $n_p$ whereas 
for weak  differential rotation  ${\cal R}$, it decreases with $n_p$.
Fig.\ref{KIC444}(right) displays the observed splittings derived by Di Mauro et al. (2018) 
as well as those provided by my long-time colleague B. Mosser (Mosser, 2019, priv. comm.) 
for the star KIC4448777 discussed in the previous section. The splittings are
normalized to the constant value $0.4 ~\mu Hz$  (close to $\delta \nu_{max}\sim \Omega_{core}/4\pi$) as a function of 
the radial order $n_p =\nu/\Delta \nu$. The  pattern of splitting variation with frequency as obtained using  an asymptotic fit
  to the observed values is also shown. These observations are compared with
 the theoretical  curve $\delta \nu_{max}/0.4~\mu Hz = 1/(1+q {\cal N}) $  which  
reproduces well the observed slight decrease of   $\delta \nu_{max}$ with frequency. 
Comparison with  the theoretical ratio  $$ \frac{\delta \nu_{min}}{0.4 \mu Hz} \sim  \frac{1+2 {\cal R}/(q{\cal N})} {1+1/(q {\cal N})} $$  
shows that it captures well the behavior of the local splitting minima
as a function of $n_p$. It is plotted 
for various values of the rotation gradient ${\cal R}$. 
Below  ${\cal R}< 0.01$, the curves are superimposed, showing that  no information can be obtained
when the rotation gradient is too small. On the other extreme,   values of ${\cal R} $ larger than 0.25 are clearly in
 disagreement with the observations for that star. A much more precise determination of the local minima appears 
to be necessary in order to obtain a real constraint on the location of the differential rotation. 
 A more detailed study involving a large sample of RBG stars is in progress (Goupil and Mosser, 2020, in prep.).

\paragraph{Period spacing} 

\begin{figure}[t]
 \centering
 \includegraphics[width=8cm,clip]{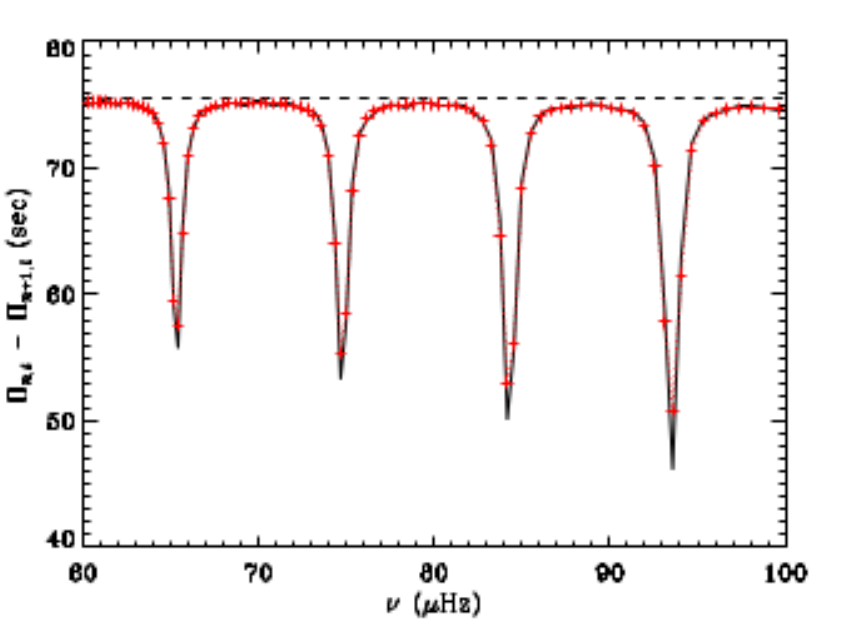}       
   \caption{\label{fig11} \small \it  Numerical period spacings computed for $\ell=1$ modes for a 1.3$M_\odot$
 red giant model as a function of frequency (the discrete values are continuously connected with the black solid lines).   The dashed horizotal line  
  represents the asymptotic period spacing. The red lines and symbols show the period spacings computed using by Eq.\ref{dPdzeta} 
(credit Hekker et al., 2017).}
 \end{figure}

The period spacings $\Delta P_n = P_n-P_{n-1}$ numerically  computed from a stellar model of an evolved  1.3 $M_\odot$ star 
 are plotted as a function of the frequency in  Fig.\ref{fig11}.
The same recurrent pattern is observed than for the inertia ratio  and the rotational splittings. Actually 
deriving Eq.\ref{cond1} with respect to the radial order, Mosser et al. (2015) found that  
\begin{equation}\label{dPdzeta}
\frac{\Delta P}{\Delta \Pi_1} \sim \zeta(\nu) 
\end{equation}
 The validity of this relation can be established  by comparing the   numerical values of $\Delta P$  for a stellar model of red giant 
on one hand and $\zeta (\nu)~ \Delta \Pi_1$  on the other hand.  Fig. \ref{fig11} 
shows a perfect agreement when $\zeta$ is also
 computed numerically. When the asymptotic expression for $\zeta$ (Eq.\ref{zeta2}) is used, some deviations can occur
 because of some departures  from the  asymptotics. 

With $\zeta(\nu)$ given by Eq.\ref{zeta2},  one obtains
$$\frac{\Delta P}{\Delta \Pi_1} =  
 \Bigl(1+ \frac{1}{q} ~ \frac{\cos^2  \theta_g }{\cos^2 \theta_p }~ \frac{\Delta \Pi_1 ~ \nu^2}{\Delta \nu}  \Bigr)^{-1} $$

An alternative but equivalent expression  can be derived from Eq.\ref{cond2} (Jiang \& Christensen-Dalsgaard,  2014, Cunha et al.,  2015). 
Eq.\ref{cond3} is rewritten as 
$$ \frac{\omega_g}{\omega} +\phi = \pi (n+1/2)$$
from which we compute 
$$\omega_g ~\Bigl(P_{n+1} - P_n\Bigr)  +\phi_{n+1}-\phi_n= \pi$$ 
and derive
$$   \omega_g ~ \Delta P  \Bigl( 1 +\frac{1}{\omega_g} \frac{\Delta \phi}{\Delta \omega} \frac{\Delta \omega}{\Delta P}\Bigr) =\pi$$
which can easily be rewritten as: 
$$    \frac{\Delta P}{\Delta \Pi_1} =  \Bigl( 1 -\frac{\omega^2}{ 2\pi  \omega_g} \frac{d \phi}{d \omega}  \Bigr)^{-1}$$

From the expression of $\phi$, we obtain 
 
$$\frac{d\phi}{d\omega}  = \frac{q}{ \sin^2\theta_p+ q^2 ~ \cos^2 \theta_p}    \frac{d\theta_p}{d\omega}  
 = \frac{q}{ \sin^2\theta_p+ q^2 \cos^2 \theta_p}    \frac{1}{\omega_p}$$
then
$$    \frac{\Delta P}{\Delta \Pi_1} =  \Bigl( 1 -\frac{\omega^2}{ 2\pi  \omega_g  \omega_p}  \frac{q}{ \sin^2\theta_p+ q^2 ~ \cos^2 \theta_p} 
     \Bigr)^{-1} = \Bigl( 1 -\frac{\nu^2}{ 2\pi  \nu_g  \nu_p} \frac{1}{q\cos^2 \theta_p} \cos^2 \theta_g       \Bigr)^{-1}
$$
where we have used  $  d\theta_p/d\omega \approx   1/\omega_p $ and Eq.\ref{cond1}.

 The highly precised $\Delta \Pi_1$ values  led to tight constraints about the 
 $N$ profile averaged over the core  hence  on the physical mechanisms that shape $N$  in the central region as discussed  in the next section.

\section{Mixing beyond the convective core of red clump stars}

A major difficulty  in the modelling of the central regions of  red clump stars is  the determination of the  location of the convective core boundary
  and of the properties (thermal and chemical stratifications) of the medium in its vicinity. 
Stellar evolution codes   use  a local  prescription 
  to locate  convectively unstable regions (basically the Schwarzschild criterion in the mixing length theory (MLT) framework
 or some improved variant formulation).  The location  usually is established by the change of sign of the 
temperature gradient difference $\nabla_{rad} - \nabla_{ad}$ : convectively  stable regions  verify $\nabla_{rad} < \nabla_{ad}$ whereas 
they are convectively unstable otherwise.  Here  $\nabla_{rad}$  represents the temperature gradient as if all the flux were transported 
by radiation and     $\nabla_{ad}$ is the temperature gradient of an adiabatically stratified fluid. For convenience,  the temperature
gradient is defined with respect to the pressure $P$ i.e. $\nabla = d\log T / d\log P$.
The central region identified as convectively unstable defines    the formal convective core  (hereafter FCC).
 This region is then assumed  instantaneously  chemically mixed and its thermal stratification assumed adiabatic.    
Generally speaking however,  one  expects that the fluid overshoots  beyond the   FCC into adjacent  
convectively stable regions where  the fluid motion is braked  down to a stop over some penetration distance.
In that region,  partial to full chemical mixing must occur and the thermal stratification can be modified.
 This then rises several questions:

\begin{itemize}

\item [-] how setting up properly the convective core boundary and its evolution with time? 

\item [-] how far does the extra -mixing region extend? 

\item [-] does the thermal  stratification become adiabatic beyond the FCC limit? 

\item [-] is the chemical composition fully or partially mixed in the penetration region ? 

\item [-]  in case of partial mixing,  what is the profile chemical gradient? How does it evolve? 
 
\end{itemize}

The evolution of the central layers of stars with a convective core 
 has been a longstanding problem over  several decades (see the review by Salaris \& Cassisi, 2017,  see also  Noels et al., 2010,
 Arnett et al. 2015, 2018,  Paxton et al. 2018 and references therein). 
Various  options for describing the region above the FCC can be found in the litterature: overshooting with instantaneous mixing
 or   overshoot with time-dependent   diffusive mixing. According to Zahn's 1991 terminology, those options refer  to the 
 case when the fluid  moves  into the stable layers but does not modify
the thermal stratification, hence the temperature gradient remains radiative. Convective penetration instead
 refers to the case when energy transfer is  efficient enough  that
  the thermal stratification becomes  adiabatic in the overshoot region.
In addition, in a convectively unstable layer according to the Schwarzschilds's criterion 
where a chemical composition gradient exists and stabilizes 
the medium (Ledoux's criterion), mixing refered to as 
semiconvection can occur (Noels, 2013). The consequences of these various options translate into 
 large uncertainties  for the further evolution and structural changes of 
 such stars.

 \begin{figure}[t]
 \centering
 \includegraphics[width=0.60\textwidth,clip]{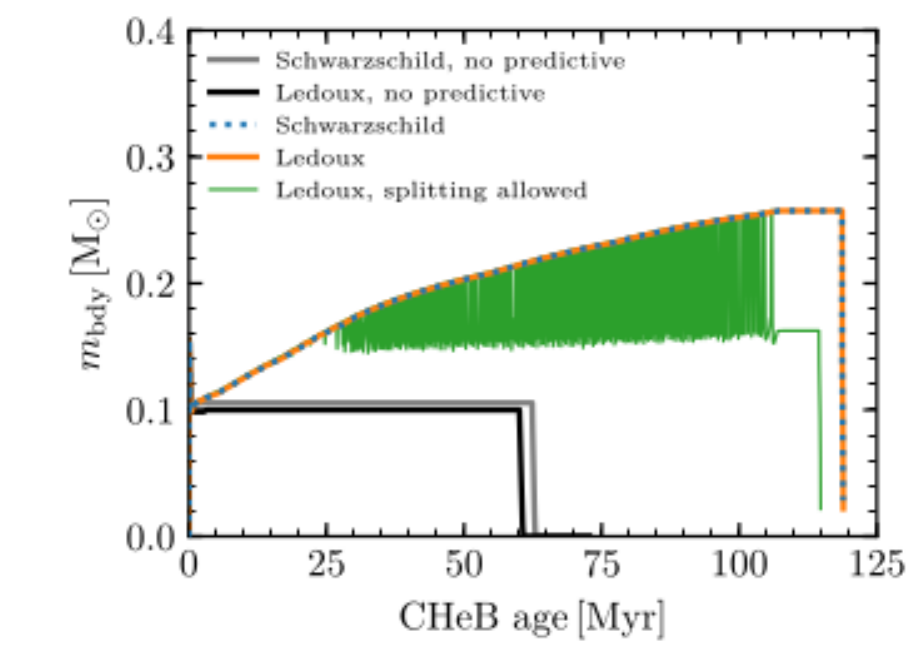}       
  \caption{\label{fig12}    \it Evolution of the mass of the  convective  core  of a  1  $M_\odot$ model with various mixing and numerical scheme options.
The FCC core (grey line) for instance does not grow in mass with time
    (credit Paxton et al, 2019).
}
 \end{figure}

A first clarification was given  by Gabriel et al. (2014)  based on earlier works. The convective boundary must be located
   at the layer where the fluid velocity vanishes  in the framework of the MLT,  the location
 of convective boundaries must be strictly set at 
  the  layer where  $\nabla_{rad} =\nabla_{ad}$  (convective neutrality). 
 Its numerical implementation must be established from the convective side of the region, this is particularly crucial 
  when a chemical  discontinuity and/or gradient exists across the convective boundary. This is  the case for red clump stars. 
Search for convective neutrality in presence of a discontinuous chemical composition across the convective core boundary
 and/or a gradient of chemical composition  in the adjacent radiative region  automatically leads to include an instantaneous mixing, mimicking fully 
mixed convective penetration (independently of the physical origin of the mixing). 
The situation is   even further complicated for red clump stars because the burning of helium into carbon and oxygen generates a local increase of the opacity 
within   the convective region which can decrease the radiative gradient there and can eventually cause  a split of the convective region
 into two   convective layers separated by a semiconvective region.    The modelling of the thermal and chemical 
stratifications  (full or partial mixing, adiabatic stratification or radiative one) in such regions therefore  remains highly uncertain.
In any case,  a proper implementation of the Schwarzschild criterion  and all formulations with extra-mixing 
lead to  larger (more massive)   convective core than FCC.
  Depending on the values of the adopted parameters  in these empirical formulations and on the adopted formulation itself, 
the convective cores  have  
   different sizes  and   different chemical profiles develop in the adjacent overlying regions (Fig.\ref{fig12}). 
   Several numerical  schema are investigated to overcome this issue (Paxton et al., 2019 and references therein).

 \begin{figure}[t]
\centering
\includegraphics[width=8cm]{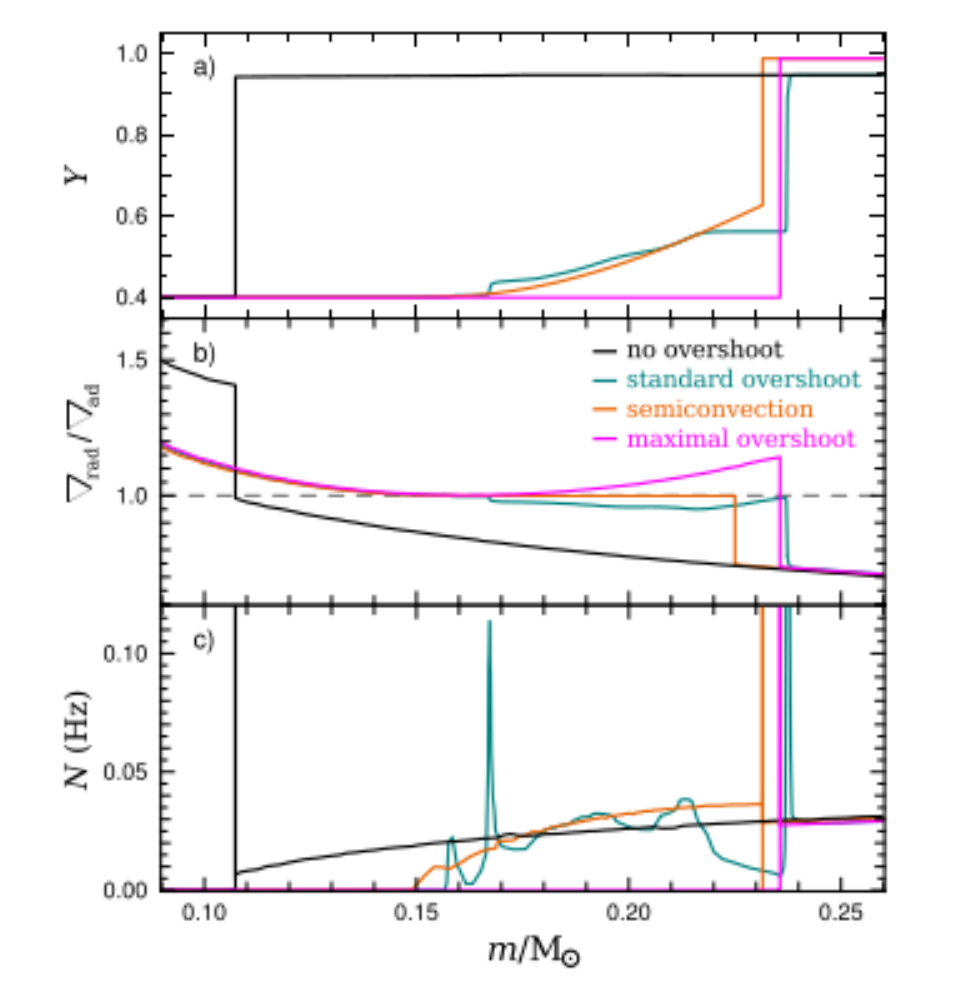}
\caption{\label{fig13}  \it \small Helium profile (top), $\nabla_{rad}/\nabla_{ad}$ profile (middle) 
and $N$ profile (bottom) for  a 1.5 $M_\odot$ model at the evolutionary phase where 60\% of helium was burned.
Four different schemes for core overshoot in a CheB star give rise to four different profiles of $N$ (Credit Constantino et al., 2015).}
\end{figure}
 
Clearly, observational constraints about the central properties of these stars can help  making  significant advances in that field.
 Montalban et al. (2013) and  Montalban \& Noels (2014) indeed  showed that the period spacing Eq.\ref{dPi1} is a sensitive
 probe of  the   properties in the central part of red clump stars.  Fig.\ref{fig13}- taken from 
  Constantino   et al. (2015)-  illustrates the  impact on the Br\"unt -Vaissala profile of various extra-mixing  schemes. 
A modification of the Br\"unt-Vaissala profile has a direct impact on the period spacing (Eq.\ref{dPi1}) and therefore offers a possibility to be 
detected observationnally.  In Fig.\ref{fig6} the increase of the mean period spacing when including 
overshooting compared to assuming a mere FCC is clearly seen.   
 
\begin{figure}[t]
 \centering
\includegraphics[width=8cm]{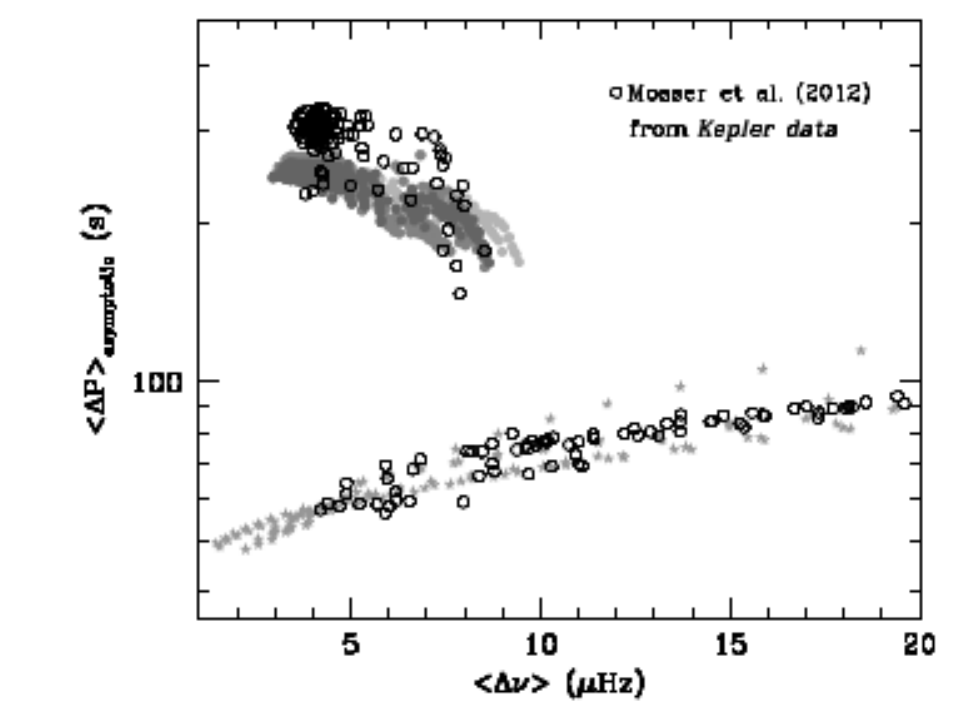}
\caption{ \label{fig14} \it \small Period spacings as a a function of  frequency  for a model of red clump (top, left samples) 
and RGB stars (bottom samples). Filled dots: asymptotic period spacings  from stellar models;  open dots: observed period spacings
 from Mosser et al. (2012a) (credit Montalban et al. 2013)}. 
 \end{figure}

The seismic observations  of red giants  and more specifically red clump stars provided by the space mission CoRoT and Kepler
 are in that sense invaluable.   Montalban et al. (2013)  compared in  a $\Delta \Pi_1- \Delta \nu $ diagram the observed period spacings of red giant 
 stars derived by Mosser et al. (2012a) 
to the theoretical  asymptotic counterpart $\Delta \Pi_1$ computed from stellar models.   
The comparison requires some care but the authors found  that stellar models of red clump stars built with the FCC option  
  provide  period spacings significantly smaller (i.e. by about 20 \%) than  the observed ones (Fig.\ref{fig14}). 
The authors  showed that adding an extra mixing (overshoot from He burning core) can help decreasing the discrepancy as it 
pushes outward the inner boundary of the g-mode cavity hence increases $\Delta \Pi_{obs}$ by 20 \% (a discrepancy of the order of  30 s).
   For instance the period spacing increases from    $\Delta \Pi  = 255~ s $ up to $\Delta \Pi  =305 ~s$  for  a stellar model  
 with $ 1.5 M_\odot$  in the middle of the He burning phase.

\begin{figure}[h]
\begin{minipage}{8cm}
\centering
\includegraphics[width=8cm]{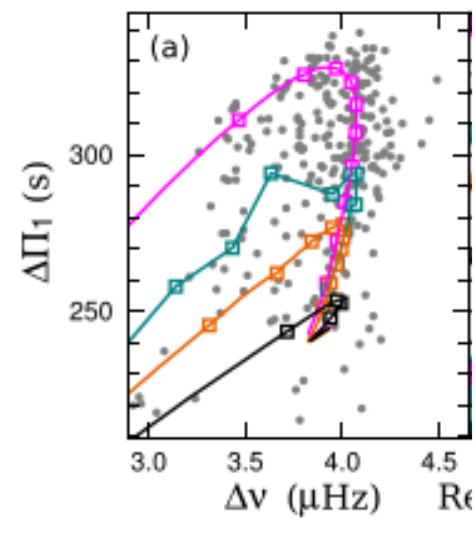}
\caption{\label{fig15} \it \small Period spacing as a a function of the large separation for stars in the red clump.
 Grey dots : observations from Mosser et al. (2012, 2014). Color lines corresponds to the evolution of a 1 $M_\odot$ 
 stellar models with several options for the modelling of extra mixing beyond the FCC (same color  than in Fig.\ref{fig13})
 {\it (Constantino  et al. 2015)}. }
\end{minipage}
\hfill
\begin{minipage}{8cm}
\centering
\includegraphics[width=8cm]{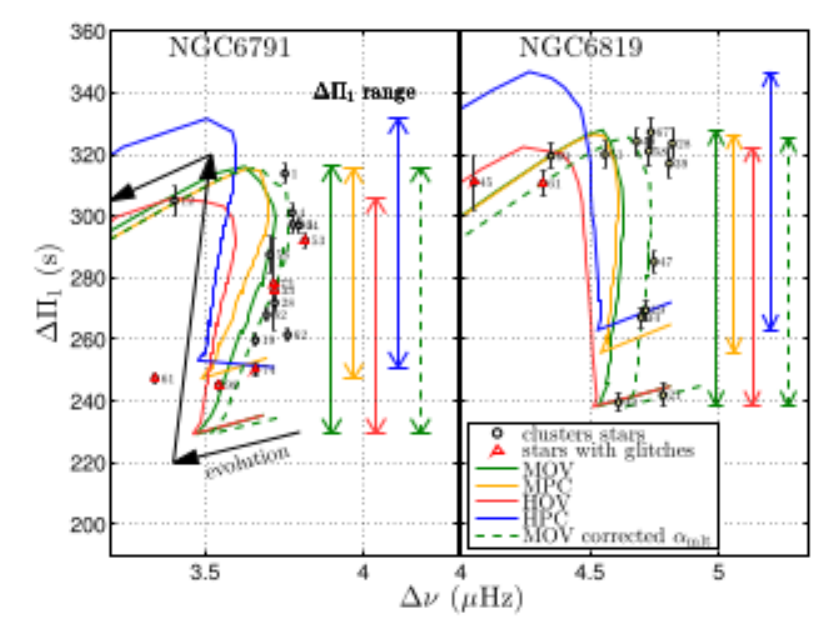}
\caption{ \label{fig20} \it \small  Period spacing as a 
function of the large separation for  red clump stars in two open clusters (dots). Open symbols connected with lines
 correspond to period spacings for stellar models  with mild overshoot (green), mild convective penetration (orange),  high overshoot (red), 
blue (high convective penetration). The arrows delimite the range $\Delta \Pi_{1, max}-\Delta \Pi_{1, min}$
 {\it (Bossini et al. 2017).}    }
\end{minipage}
\end{figure}

   Bossini et al. (2015) and Constantino  et al. (2015) further tested the impact of several options 
of mixing beyond the FCC using the period spacings derived for a set of Kepler red clump  stars  (Mosser 2012, 2014). 
Fig.\ref{fig15} taken from  Constantino et al. (2015)  clearly indicates that without
any extra mixing beyond the FCC, the period  spacings never reach  the observed maximal values
  in a $\Delta \Pi_1 - \Delta \nu$ diagram. 
They agree with  the conclusion that the chemically mixed core  must be significantly larger than the FCC core as
 it can be seen in Fig.\ref{fig15}   for instance. This confirms results from 
 theoretical studies  and numerical simulations and confrontation using classical (non-seismic) observations.

Interpretation of the $\Delta \Pi_1 - \Delta \nu$ for field red clump stars  may be complicated by their somewhat different metallicities. 
In order to avoid this  complication, Bossini et al. (2017) studied red clumps stars belonging to
two open clusters NGC 6791 (with representative  stellar models $1.15 M_\odot$ with initial $Z = 0.035$ and $Y = 0.3$ )
and NGC 6819 (with   representative stellar models $M = 1.60 M_\odot$ with initial
$Z = 0.0176$, and $Y = 0.267$).  The  seismic data were obtained by Vrard et al.  (2016).   
The authors compared  the behavior of $\Delta \Pi_1$ as a function of 
$ \Delta \nu$  for stellar models including different modelling of the regions  beyond the FCC.  
 The results are summarized in Fig.\ref{fig20}  (Fig.1 in Bossini et al., 2017). 
  At a given age and metallicity, the convective core evolution 
causes  an increase of  $\Delta \Pi_1 $ with $ \Delta \nu$. The minimal value of $\Delta \Pi_1$ corresponds to 
less massive, young stars (beginning of the CheB phase) 
while the maximal value corresponds to more massive- hence slightly more evolved- stars (end of CheB phase). 
The comparison  tends to favor the scheme where the thermal stratification in the extra-mixing region
 remains radiative at least at the beginning of the CHeB phase.    The other option (PC) seems more in agreement 
with the three low mass stars in NGC6819. The result in that last case  remains doubtful because 
 the   stars are more massive and the correct period spacings are more difficult to determine.  
For the late CHeb  phase  (most massive stars in the samples), the convection tends to split  as discussed above. 
Therefore the value of the period spacing strongly  depends 
on the adopted scheme  for the region in between the two  split parts  but in any case,  
the assumption of the largest cores seem to be ruled out.

These interesting studies show that  advances must  come from  larger samples of stars with given mass and metallicity. On the theoretical side, 
improvement of modelling during the late part of the CheB phase when  the helium content in the burning region gets small 
- and the size of the core crucially
 depends on the  modelling of the extra mixing beyond the FCC-
are particularly necessary  (Paxton et al., 2018, 2019 and references therein, Spruit, 2015, Constantino et al., 2017).

\section{Angular momentum transport in low mass post main sequence stars  }

\subsection{Rotation profiles of red giant stars}

In 1D stellar modelling, the  convective envelope of  cool evolved  stars is usually assumed to rotate  uniformly 
\footnote{although this may not be the reality, see   Brun \& Palacios 2009}. 
  If local conservation of angular momentum is maintained within the adjacent radiative region below, 
the core spins up and the envelope decelerates due to the contraction of the core and expansion of the envelope 
during the subgiant and giant phases. In that framework,  cool,  evolved stars are  expected to 
develope a strong rotation gradient between the fastly rotating core and the slowly rotating envelope. 
  
On the observational side,  a lower limit of the surface  rotation rate for subgiant and red giant stars, $ \Omega_{surf}$, can  
 be derived   from spectroscopic measurements of the projected surface rotation  $v ~\sin i$. 
The surface rotation period can also be obtained from photometric measurements of spot modulations when 
the star undergoes some type of surface activity. However because the surfaces of  subgiants and red giants are 
mostly rotating slowly, the measurements  are only available   for a small subsample of stars biased 
toward the rapid rotators (Tayar et al., 2015; Ceillier et al., 2017).
On the other hand, seismic measurements of the mean  rotation rates of the core of evolved stars, $ \Omega_{core}$,  
can be obtained from the  maximum value 
reached by the rotational splittings $\delta \nu_{max}$
 (Eq.\ref{split})  or from inversion  methods using individual splittings  (Fig.\ref{fig21}).  
They are found of the order of  a few ten days for subgiants  and RGB stars and of some hundreds days
 for red clump stars   (Beck et al., 2012; Mosser et al., 2012b, Deheuvels et al., 2012, 2014, 2015, Di Mauro et al.  2016, 2018).  
They  actually are in good agreement with the rotation
of the  hot B subdwarf (sdB) stars which might be their descendants (Charpinet et al., 2018).
 The observed  core rotation rates of subgiant and red giant stars are however much smaller 
than  what can be expected when assuming instantaneous angular momentum (hereafter AM) transport in convective regions and 
 local conservation of angular momentum  in radiative regions

\begin{figure}[t]
 \centering
 \includegraphics[width=8.cm,clip]{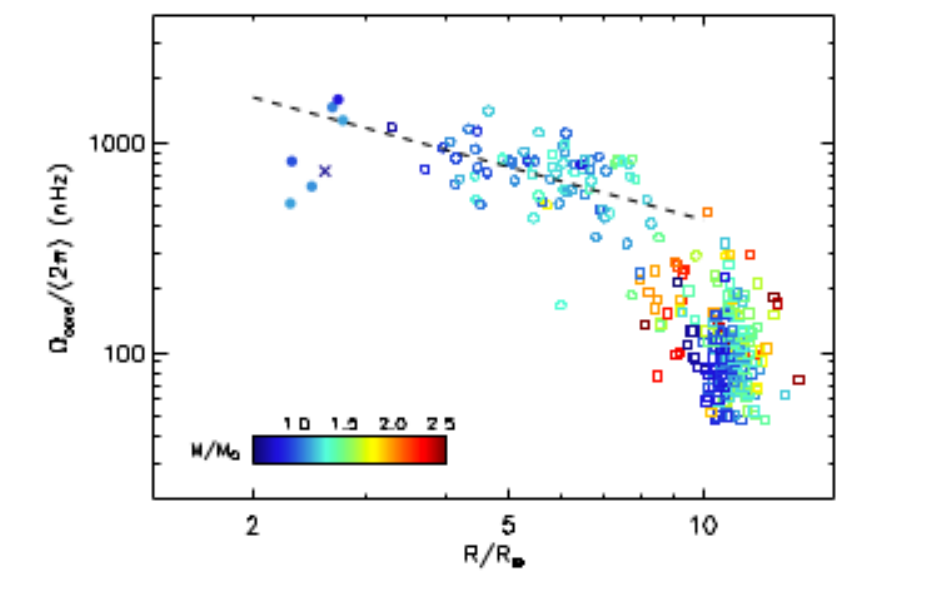}        
 \includegraphics[width=8.cm,clip]{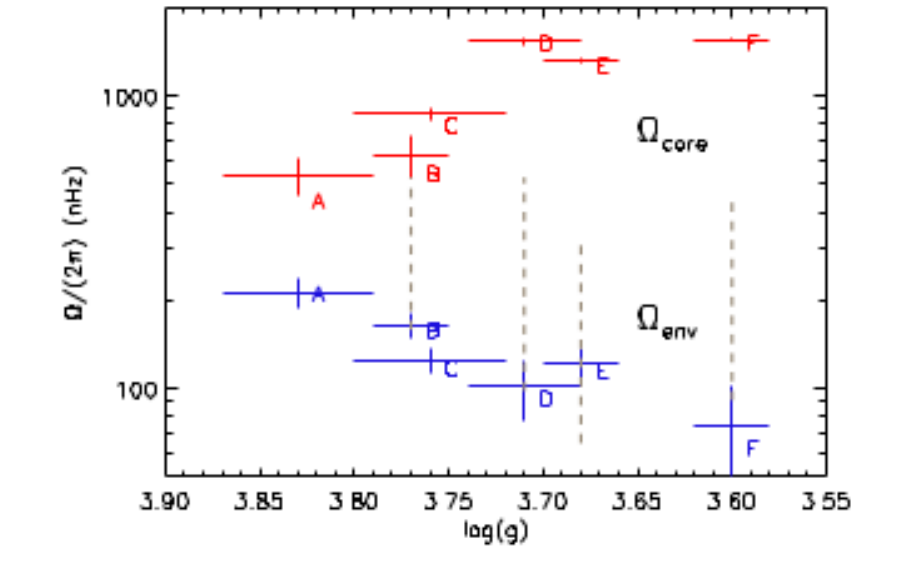}        
   \caption{\label{fig21a}  \small \it   
 left: seismically measured core rotation of subgiants and red giants as a function
 of the seismically  determined stellar radius. Evolution goes with increasing stellar radius. Stellar radii  
in the range 2-8 $R_\odot$ correspond to   RGB stars. 
Clump stars have stellar radii larger than  8 $R_\odot$. 
 Right: Core and envelope seismic rotation rates for six subgiant stars and one red giant (Credit Deheuvels et al., 2014). }
 \end{figure}

\begin{figure}[t]
 \centering
 \includegraphics[width=18cm]{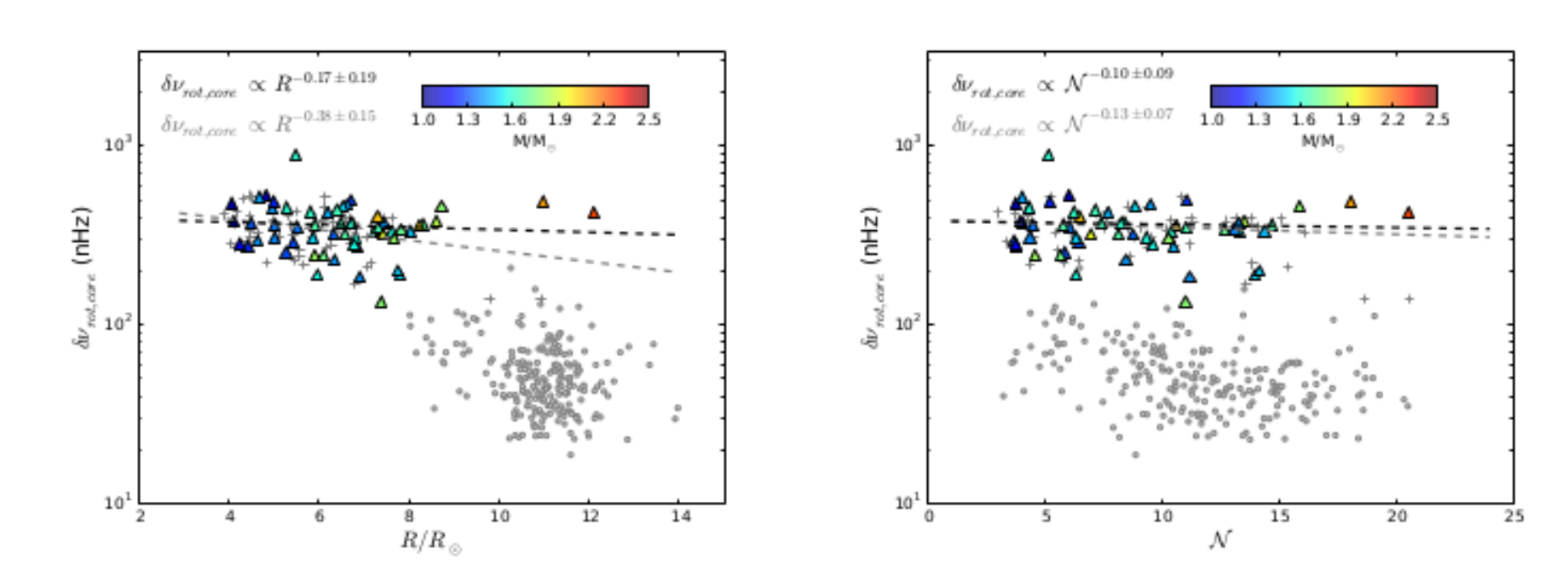}        
   \caption{\label{fig33}  \small \it  left: core contribution to the  rotational splitting
as a function of the stellar radius used as a proxy for age.
 Coloured symbols are for data from Gehan et al. (2008). 
Mosser et al(2012b)'s measurements on the RGB and on the clump are represented by grey crosses and dots, respectively.
 Right:  same as left but with the mode density ${\cal N}$ as a proxy for
 age  (Credit Gehan et al., 2018). }
 \end{figure}

Further, individual studies for six subgiants indicate  that the star core 
rotation accelerates when  evolving  toward the RGB (Deheuvels et al., 2014) whereas  a statistical study of 300 giants including 85 RGB stars 
showed that the core rotation of RGB stars  seems to slow down when the star ascends the RGB 
(Mosser et al.,  2012b).  This is summarized in Fig.\ref{fig21a} where the core rotation 
is plotted as a function of the radius (proxy for evolution). Such evolution of the core rotation cannot be reproduced by 
standard evolution models which do not include  an efficient AM 
transport in the radiative interior of  these cool stars. 
However the above view has been recently challenged :  are  the RGB cores really slowing down when ascending the giant branch? 
With a more sophisticated data analysis process,  Gehan et al. (2018) obtained the core rotation rates for a much larger  sample of red giant  stars.  
The stellar radius is usually used as a proxy for the age of the star on the ascending red giant  branch. Fig.\ref{fig33}  reveals 
that despite a high dispersion, the spin down of the core rotation rate of the RGB stars 
  seems to be weaker than previously established by Mosser et al. (2012). Further,  Gehan et al. (2018) emphasized that  
the star does not enter the RGB phase with the same radius depending on its mass. The authors then argue that 
a better age indicator is the averaged number of  g-dominated modes existing between two p-dominated modes, ${\cal N}$. With that indicator as a proxy for the age,
the core  does not seem to slow down when the stars ascends the giant branch  (Fig.\ref{fig33}) 
and this is found  to hold  true independenly of the mass of the star.

An additional  constraint is given by the seismic measurements  of the core-to-envelope  rotation ratio 
 $\Omega_{core}/\Omega_{env}  $ which  can  
 be obtained in favorable cases as mentionned in Sect.2.
Fig.\ref{fig21a}, right, illustrates the
case when  both envelope and core rotation rates are seismically determined  for six subgiant  stars (Deheuvels et al., 2014).
 Theses estimates indicate that the ratio  $\Omega_{core}/\Omega_{env}  $ lies  in the range 2.5-20 and increases with evolution.
For more evolved red giants, the contribution from the envelope to the rotational splittings becomes negligible and reliable estimates of  
 $\Omega_{env}$ can no longer be obtained from the linear  relation  Eq.\ref{deltanu}.

\begin{figure}[t]
 \centering
 \includegraphics[width=7cm,clip]{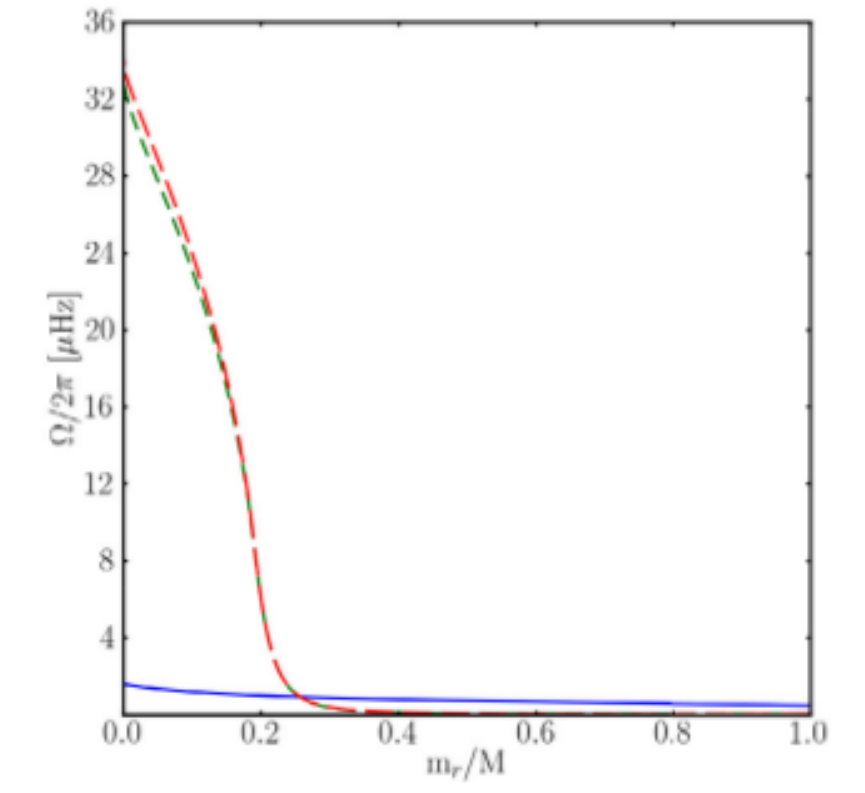}        
 \includegraphics[width=8.5cm,clip]{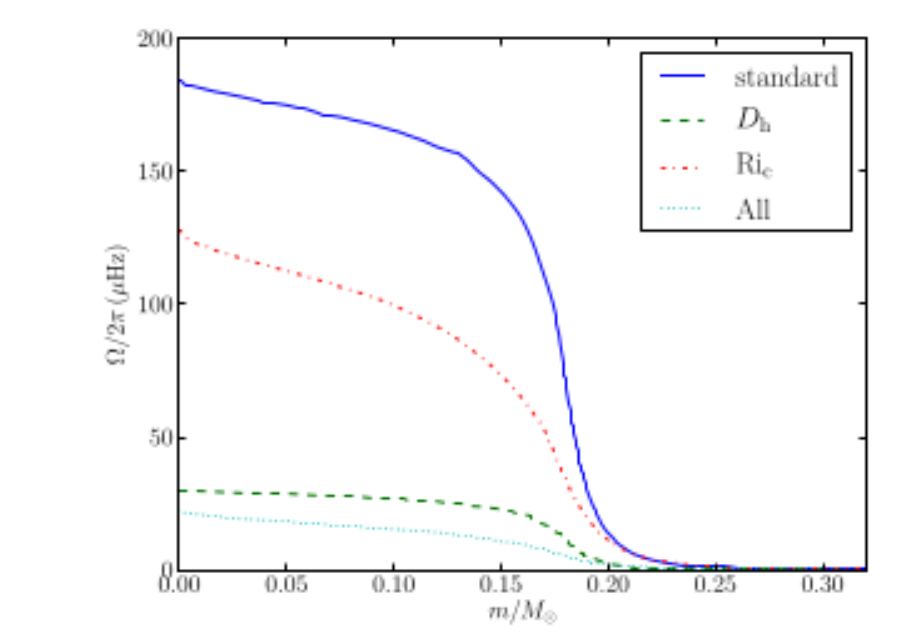}        
   \caption{\label{fig22} \small \it Left:  Rotation profiles for  0.84 $M_\odot$  models of the early red giant 
star KIC 7341231 at an age of 13.40 Gyr obtained with several assumptions 
about AM transport in the  radiative  region (black: assuming shellular rotation AM transport;
  blue : assuming  uniform rotation on the main sequence, red : conservation of angular momentum during the subgiant phase : same as blue but $X_c=0$).
 The observed rotation range is   $710 \pm 51$ nHz (Deheuvelset al., 2012) at the very bottom of the plot   
    (Credit Ceillier et al., 2013). Right: Rotation profiles for 1.3 $M_\odot$  models   at the base of the RGB when $R = 3.73 R_\odot $ 
calculated assuming standard viscosity coefficients (label 'standard'), 
a vertical turbulent viscosity $D_V$ computed with the Richardson number $Ri_c = 1$ ($Ri_c$) , a horizontal viscosity coefficient 100 times
 the standard value ($D_h$ ), and  combining all (All) (Credit Marques et al., 2013}.
 \end{figure}

The observations  thus reveal that the inner cores of the PoMS stars rotate significantly faster than
the envelopes albeit much less faster than predicted by standard stellar models when local conservation of AM is assumed in radiative regions. 
 One therefore must admit  that an efficient  AM transfer is at work from the inner to the outer 
layers of these stars to compensate for the core spin up due to the contraction of 
the central layers beyond the MS phase.  So the issues are:
 
- what mechanism(s)  are able to slow down  the core  of red giants with evolution and reduce(s) the the radial differential rotation of 
 evolved stars compared to what would  be expected from evolution with local conservation of AM ?

- if several processes  are at play, what is the dominant one?

- can these mechanisms be modelled properly enough in 1D stellar models to account for the above observations?

\subsection{What mechanism(s) can slow down enough the core of post-main-sequence  stars ?}

Several mechanisms of AM transport in radiative interiors  can be invoked  (Talon, 2008; Maeder 2009).

\paragraph{AM transport by hydrodynamical processes:}

At the basic level, structural changes due to evolution naturally generate differential rotation in radiative regions. This rotational shear can 
 drive hydrodynamical instabilities  leading to turbulence and therefore turbulent AM transport (Talon 2008).
Historically the AM transport in radiative zones of  1D stellar models was then modelled as a pure diffusive process originating from 
 the turbulence induced by a rotational shear. In that case,   the AM vertical  turbulent diffusion coefficient is
parametrized and calibrated on observations (Endal \& Sofia 1978, 1981; Pinsonneault et al. (1989);  this is still the case in several evolutionary codes.  
However with studies  started more than 20 years ago, Zahn (1992) and co-workers  
   showed    that AM  can be advected by large scale   meridional motions in radiative regions of rotating stars, enhancing the rotational shear 
and competing with turbulent diffusive AM transport.   This circulation is sustained  by  structural changes due to evolution,  turbulence  and    by   AM loss at the surface for low mass stars with convective envelopes. 
The rotation profile  in radiative   stellar regions of rotating stars  therefore results  from an advective-diffusive AM  
 transport   on secular time-scales    (Zahn, 1992; Talon \& Zahn, 1997; Maeder \& Zahn 1998;     Mathis \& Zahn 2004; Zahn 2013;  
  see Palacios 2013 for a review).

While  the assumption of a rotation rate  a function of radius only, $\Omega = \Omega (r)$,  was considered from the start in 1 D stellar modelling, 
 Zahn (1992)  justifies this so-called shellular rotation  as the result of a strongly anisotropic turbulence 
which causes a rapid homogeneization on horizontal surfaces. This assumption allows to reduce the AM transport description
to a one dimensional process that-is  in the  vertical (i.e. radial) direction.  In radiative regions,  the local specific AM, 
$j(r) = r^2 \Omega(r) $ then  obeys a time-dependent equation of the form:
\begin{equation}\label{AM}
 \frac{\partial r^2 \Omega}{\partial t} =  - \dot r ~ \frac{\partial r^2 \Omega}{\partial r} 
-\frac{1}{ \rho r^2}\frac{\partial \rho r^2  {\cal F}_{tot}  }{\partial r}  +\dot j_{wind}
\end{equation}
where the momentum flux is given by 
  \begin{equation}\label{flux}{\cal F}_{tot} = {\cal F}_{circ}+  {\cal F}_{shear}
 \end{equation} and
 \begin{equation}\label{flux1}{\cal F}_{circ}    =  -\frac{1}{5} r^2 \Omega  ~U_2  ~~~~~~~~~~; ~~~~~~~~~~~~~~
~{\cal F}_{shear}    =  -  r^2 \Omega  ~ \nu_v ~\frac{\partial \ln \Omega}{\partial r} 
\end{equation} 
where $ r(m)$ is the radius enclosing the mass, $ \dot  r  $ is  the time derivative of the radius, 
$\dot j_{wind}$ stands for the AM loss by a surface magnetized wind.
The effect of the  wind is to decelerate the rotation of the convective envelope, which is assumed to rotate rigidly at the same angular
velocity than the surface. AM loss by magnetized wind is a complex 3D process, its modelling keeps on being improved
 and its impact on  the rotation profile evolution being studied (Amard et al., 2016  and references therein).
Equations and prescriptions for the vertical meridional circulation  speed,  $U_2$,  and the vertical diffusion coefficient  
 for  vertical transport induced  by a vertical shear , $\nu_v =\nu_{shear}$,
entering   Eq.\ref{AM}-\ref{flux1} above  were established by Zahn (1992), Talon \& Zahn (1997),  Maeder \& Zahn (1998), and Mathis \& Zahn (2004). 
The above equation is coupled to the evolution equations for  the chemical elements  not given here (Talon 2008, Maeder, 2009).

\begin{figure}[t]
\begin{minipage}{8cm}
\centering
\includegraphics[width=8cm]{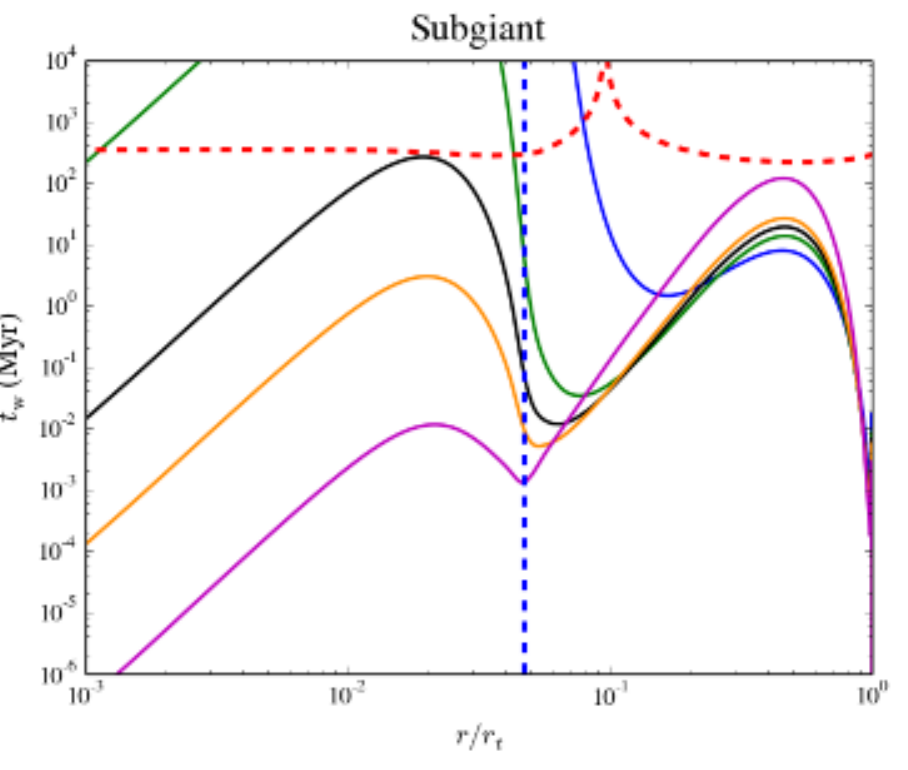}
\caption{\small \it IGW time-scales (solid lines)
 as a function of the normalized radius in the radiative zone of a 1 $M_\odot$ subgiant model. 
Colors correspond to different values for the amplitude of the differential rotation. The red and the blue dashed lines
represent the contraction or dilatation timescale and the location of the
hydrogen-burning shell, respectively. (Credit Pincon et al. 2017). \label{fig21}}
\end{minipage}
\hfill
\begin{minipage}{8cm}
\centering
\includegraphics[width=8cm]{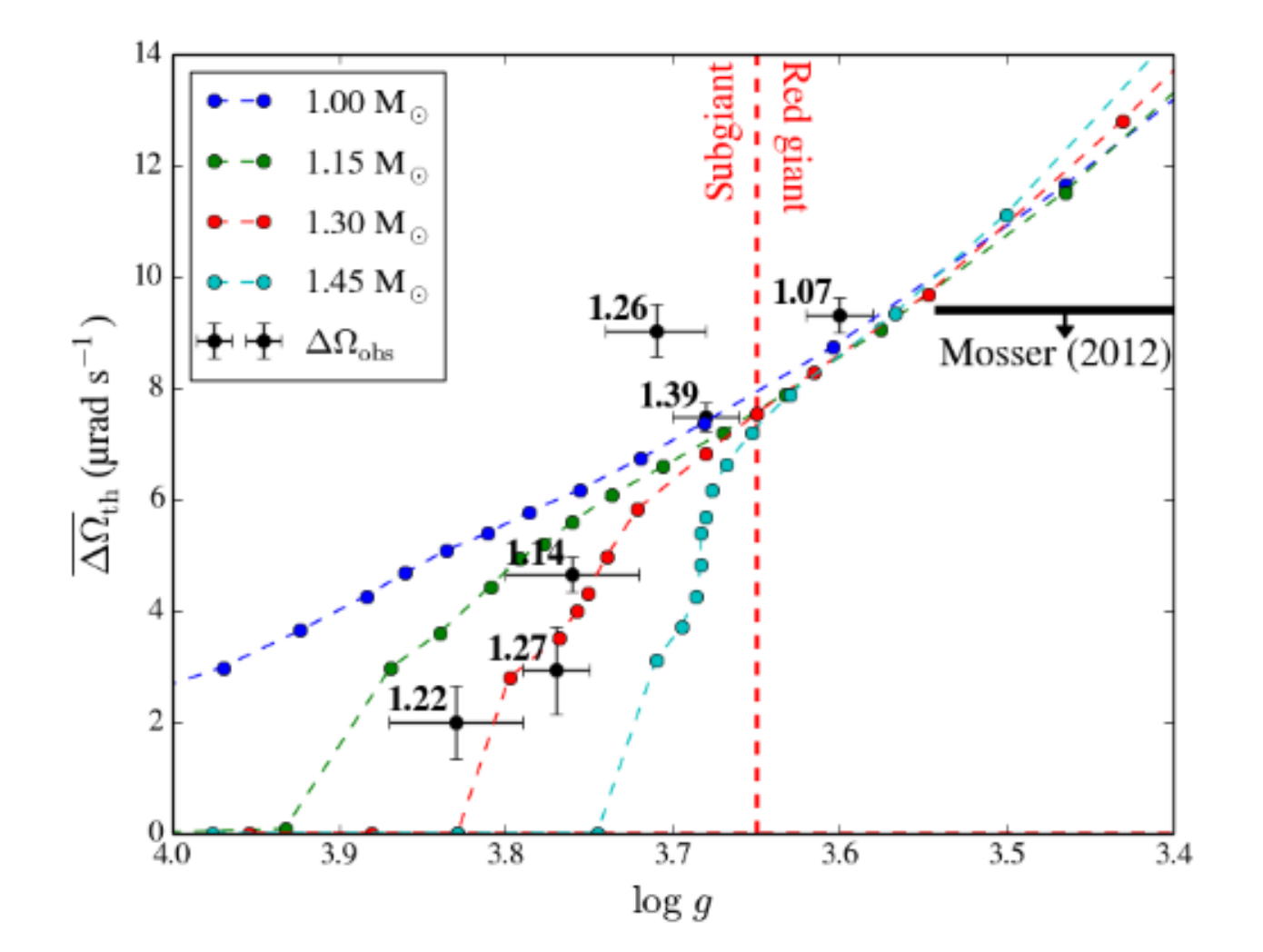}
\caption{ \small \it Evolution of the theoretical threshold for the amplitude of the radial-differential rotation $\bar{ \Delta \Omega}  $  resulting from IGW AM  transport
as a function of the stellar evolution proxy $\log g $. The vertical red dashed line marks the transition between the subgiant phase to the beginning of the RGB.
 The observed rotation contrasts  $\Omega_{core}- \Omega_{surf}$ for the  subgiants  are   derived from Deheuvels et al. 2014 and shown as error bars.
The horizontal thick black line  indicates the  observed maximum amplitude of the differential amplitude on the RGB as derived  by
 Mosser et al. (2012) (Credit Pincon et al. 2017).  \label{fig24}}
\end{minipage}
\end{figure}
 
Advective-diffusive AM transport is  implemented  nowadays in several evolution codes but as already mentionned, the resulting AM  transport 
appears to be  insufficient to slow down enough the core rotation of red giant stars (Fig.\ref{fig22}), 
  a discrepancy which reaches one to two orders of magnitude depending on the star and the physical assumptions in the AM transport along the evolution 
(Eggenberger et al. 2012;   Marques et al., 2013; Ceillier et al.,  2013).
  While the advective-diffusive approach and the shellular approximation may still be reasonable  for low mass, slowly rotating stars, 
several uncertainties remain  in the modelling of the  AM transport by hydrodynamical processes in stars. 
  In particular the turbulent diffusion coefficients entering the 
equation for AM and chemical  transports are mostly empirical  (Meynet et al., 2013)
and are currently being revised on more physical ground (Maeder et al 2013,  Mathis et al., 2018).
 
In an attempt to render count of the rotation profiles of evolved low mass stars, the attention then turned to other transport   processes  as likely  good candidates: 
AM transport by gravity waves, ${\cal F}_{waves }$
 and normal modes  ${\cal F}_{modes}$ and 
AM transport induced by magnetic instabilities.
 
\paragraph{Transport driven by internal gravity waves (IGW) ${\cal F}_{waves}$: } IGW are convectively excited in or near the base of the convective envelope
 and propagate in the radiative interior where they are eventually dissipated. 
The net AM transport  results  
from the  opposite contributions  of  retrograde and progade IGW. 
Assuming shellular rotation and core rotating faster than the enveloppe, 
the differential rotation serves as en efficient filter for the waves: only retrograde waves   can propagate  deep down toward the core 
and dissipate there  while the prograde waves are rapidly damped close to the convective bottom of the envelope
 (Press, 1981; Schatzman, 1993; Zahn et al., 1997;  for reviews, see Talon 2008;  Mathis, 2013;  Mathis \& Alfvan 2013).  
 This asymetric propagation can  decelerate  the core if it occurs on  time scales  shorter than the evolutionary time scale.
 The  IGW time scale  strongly  depends    on  the radiative dissipation. The radiative damping of the waves increases downward, therefore IGW 
must be excited to sufficient amplitudes at the bottom of the convective envelope 
in order to  reach the core and extract sufficient AM there. The  IGW time scale  therefore also  depends  
on the wave excitation process at the base of the convective envelope and is  one of the main
 uncertainties of this AM transport process.

  Two types of excitation mechanisms have so far been studied for post main sequence stars:  
the IGW excitation by turbulent convection is not efficient enough for the IGWs to overcome the strong dissipation at the H shell burning  
 neither during  the subgiant phase nor later during the giant phase. It cannot therefore explain the  observed 
subgiants and RG  core rotations (Talon \& Charbonnel, 2008; Lecoanet \& Quataert, 2013;  Fuller et al. 2014; Pincon et al., 2016).
 On the other hand, IGW excitation triggered by the penetration of turbulent plumes  overshooting the radiative region below the CZ 
generates an AM transport on times scales that can be shorter than the evolutionary time scale (Fig.\ref{fig21}, Pincon et al. 2017). 
 With this type of excitation, IGW are able to  slow down the  subgiant core rotation when the
 differential rotation between the core and the bottom 
of the envelope  is large enough  to decrease the dissipation of the retrograde waves. Pincon et al. (2017) suggest that a 
 self regulating process  imposes the rotation rate to be close to the rotation threshold which seems in agreement with observations (Fig.\ref{fig24}).
  Detailed computations of the evolution of the rotation profile of a low mass  star including 
both IGW excitation mechanisms remain to be done.  It must also be stressed that  
the IGW  generation and propagation  suffer from additional complexities such as  critical layers (Alvan et al., 2013), 
the impact of rotation (Andre et al., 2018) and magnetic field (Loi  \& Papaloizou, 2018) 
that are being investigated  in  current  theoretical and numerical works.  
The  net result of IGW  is the propagation of  successive  fronts from the center to the surface. The  internal rotation rate is then 
expected to be non uniform and the location of the rotation  maximum to change  with time (Talon \& Charbonnel, 2005; Alvan et al., 2013).
 Such a specific rotation profile, if detected, 
 would indicate that IGW are operating efficiently in evolved stars, however  the possibility of a  seismic  detection  of a IGW induced 
rotation gradient  will  depend on the  location  of the front within the star when
 we observe the star. If the front happens to be in the intermediate evanescent region of the star, it will likely not be detected. 
 
In any case, for red giant stars, the IGW damping becomes so large  that IGW seem unable on their own  
  to  decrease  the core rotation of  RGB  (Pincon et al., 2017). At this evolutionary phase, another AM transport mechanism must operate.

 \begin{figure}[t]
 \centering
  \includegraphics[width=8cm,clip]{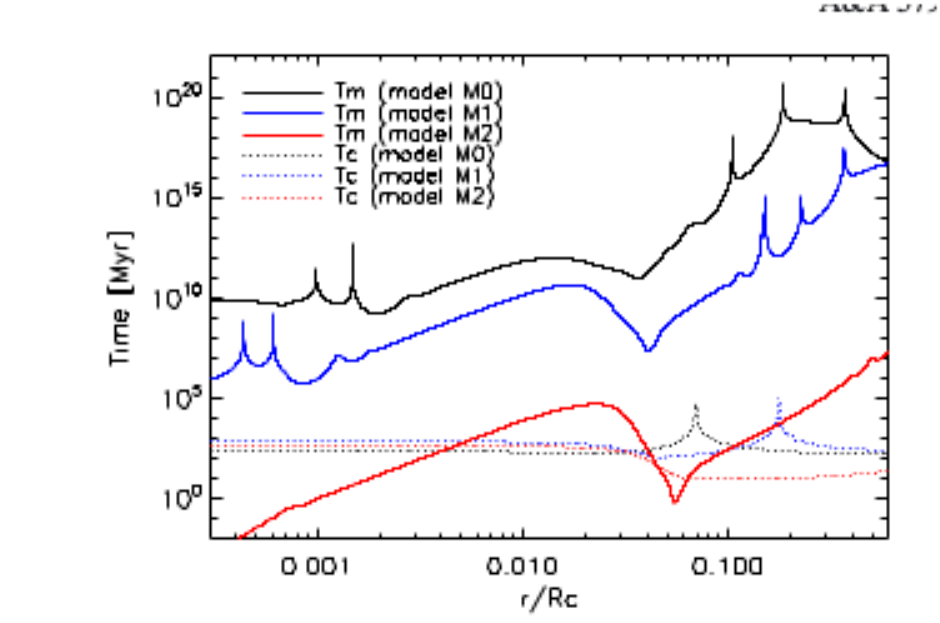}        
   \caption{\label{fig23}  \small \it  Mixed mode AM transport timescales (solid lines)  versus   radius normalised
to the radius of the base of the convective envelope for one subgiant model, one model at the bottom of the RGB and one  RGB
model located   just below the RG bump.  The dotted lines correspond to the timescale associated with the evolution of the star  
(Credit Belkacem et al 2015b). }
 \end{figure}

\paragraph{Transport by mixed modes}
  Prograde and retrograde mixed modes 
are differentially damped in   presence of rotation, allowing a net transport of angular momentum (Townsend, 2014 and references therein).
Building on that picture,  Belkacem et al. (2015a)  investigated the  AM transport by mixed modes of stars in  post main sequence phases. 
 Belkacem et al. (2015b) then 
estimated the rate of angular momentum transported
by mixed modes for  $1.3 M_\odot$ models in the subgiant and red giant phases.  The  AM flux is found to increase with 
evolution from the subgiant to the  top of the red giant branch  due to a combination of several factors. 
 The efficiency of the AM transport by modes was then  assessed by comparing  the timescale associated with   the transport
of angular momentum by mixed modes   with the time scale associated with the core contraction of the star.
 The comparison is shown in Fig.\ref{fig23} which 
clearly indicates that the AM transport  process is inefficient (too slow) for  subgiants and giants at the base of the RGB but   is able
 to counterbalance the structural changes for a giant higher up on the RGB.
 A more quantitative investigation requires  evolutionary calculations including the effect of
mixed modes together with other identified AM transports which are yet to be done. This will allow to check a posteriori 
 assumptions made in deriving the  prescription  for  AM transport by mixed modes.

\paragraph{Magneto-hydrodynamic instabilities}

\begin{figure}[t]
\begin{minipage}{8cm}
 \centering
 \includegraphics[width=9cm,clip]{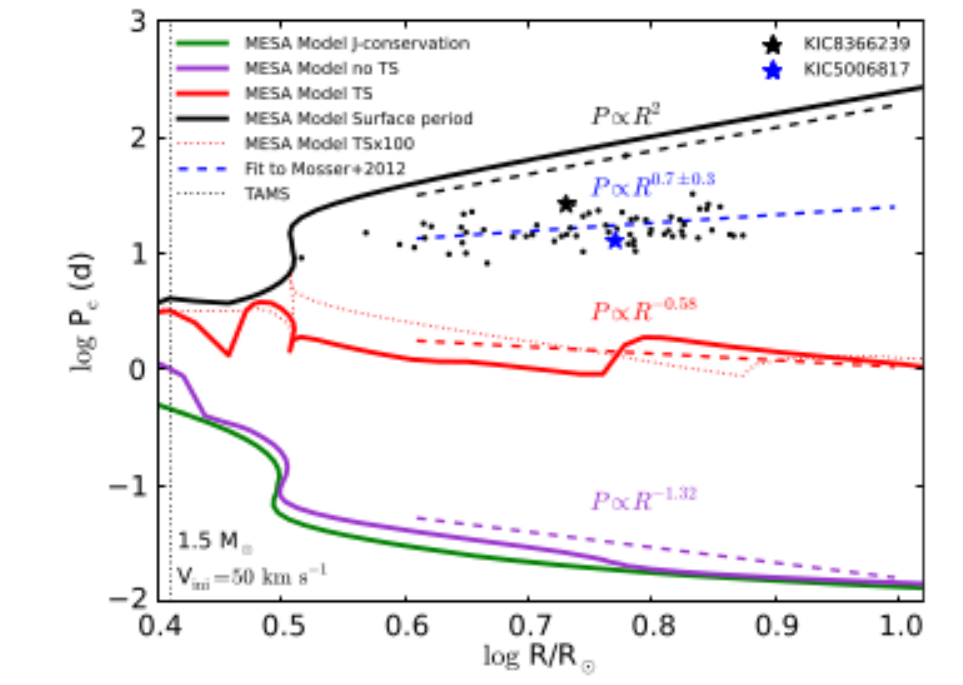}        
   \caption{\label{fig26}  \small \it   PoMS evolution of the  core rotation period 
as a function of the stellar radius for a 1.5 $M_\odot$  model
assuming different assumptions for AM transport: 
no  AM transport (green), AM  transport due  rotational instabilities (purple) and due to  magnetic torques in radiative
regions (red) and   a case when the Tayler-Spruit diffusion coefficient has been multiplied by a factor of 100 (red dotted line).
  Observations (Mosser et al., 2012, Beck et al., 2012) are represented as dots (Credit Cantiello et al., 2014).}
\end{minipage}
\hfill
\begin{minipage}{8cm}
\centering
 \includegraphics[width=9cm,clip]{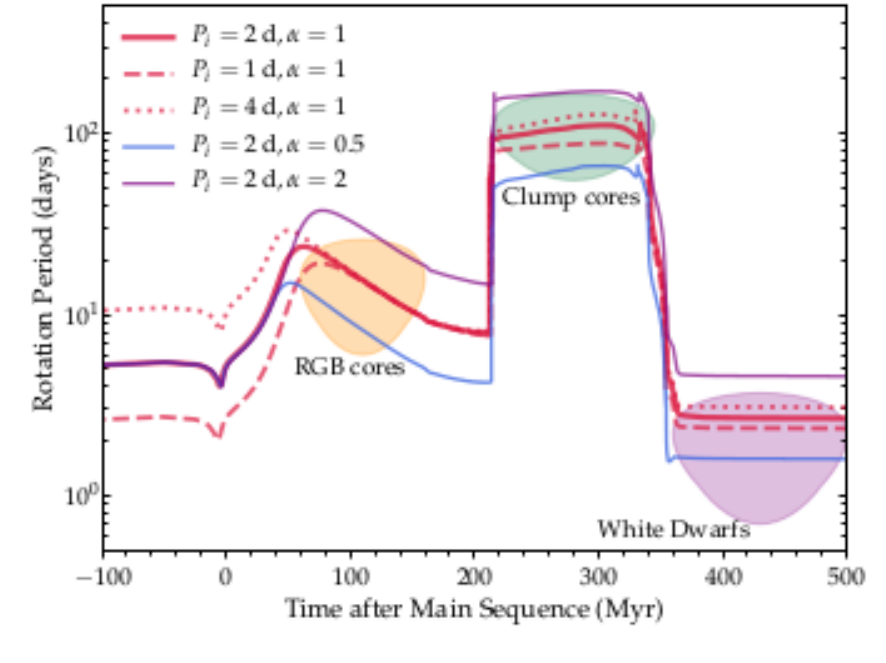}        
   \caption{\label{fig27}  \small \it PoMS evolution of the core rotation period as a function of the the time after main sequence
 for a 1.6 $M_\odot$ model 
  including  the revised  prescription for the magnetic diffusivity related to a Taylor instability.
 The core rotation rates are represented  for different values of the parameters.  The coloured areas define the observed range for  
 RGB, RG and WD phase (for the later,  Hermes et al. 2017)  (Credit Fuller et al., 2019).  
}\vskip 2.truecm
\end{minipage}
 \end{figure}

 The Tayler magnetic instability is believed   to occur in stellar interiors
 (Spruit, 1999;  Goldstein et al,  2018).  
What  is still debated is the existence of the Tayler-Spruit dynamo  mechanism. In any case, 
assuming the prescription proposed by Spruit (2002) for the AM diffusivity,  the Tayler-Spruit dynamo   was shown to be 
 not efficient enough to counterbalance the core acceleration due to its contraction with evolution for the red giant stars (Cantiello et al. 2014).
Fig.\ref{fig26} shows the rotation period as a function of the radius for the  subgiant and  stars on the red giant branch. 
These observations are compared to the evolution of the rotation period with the radius  for different types of stellar models built 
assuming different assumptions about the internal AM transport. Fig.\ref{fig26} shows that they all fail to reproduce the observations.

Recently,  Fuller et al. (2019) revisited the saturation mechanism of the Taylor instability. As a result,  the magnetic field can grow to larger
 amplitudes, leading to a larger magnetic  torque and  therefore to a larger  AM diffusivity 
$\nu_{B} \sim r^2 \Omega^3/N_{\mu}$  where $N_\mu$ is the dominant component of the Br\"unt-Vaiss\"al\"a  due to a chemical gradient. 
  The authors then showed that this  new prescription implemented in an evolutionary code  is able to reproduce the magnitude of the 
low RGB core rotation rates, clump stars and white dwarfs (Fig.\ref{fig27}).
 Prior to these stages, nearly rigid rotation in the radiative zone is maintained during and beyond the end of the MS   during the early 
subgiant  phase. However  the model predicts a spin down for the late subgiants  and a spin up evolution for the red giants when the TS instability can no longer 
 hold the acceleration due to contraction. These results are then at odds with the present observations 
that-is the core spin up of the subgiants with evolution and the core  spin down (Mosser et al 2012) or a constant spin evolution (Gehan et al. 2018) 
found for  the RGB stars. It is   not impossible that the conclusions 
 drawn from observations  continue to evolve  when a larger sample of higher quality data  become available 
as the results from Mosser et al. (2012) then  Gehan et al.  (2018) showed.
This can be particularly true for subgiant stars since the present sample is quite small.

    Another magnetic instability,   the  Magneto-Rotational Instability (MRI)  is rather triggered by a radial 
 differential rotation and can lead to AM transport (R\"udiger et al., 2014, 2015).
 Jouve et al (2015) studied  the coupled evolution of a magnetic field and differential rotation 
in an 3D  spherical unstratified radiative  shell and found that the MRI developes when the rotational shear is large enough.
The rotational shear must overcome  the stable stratification so it is not clear yet if the MRI can develope in 
 the stellar rotation/stable stratification regime. 
Spada et al.  (2016)  computed the evolution of the rotation profile of a 1.2 $M_\odot$ model including 
AM  transport in radiative interior which  assumes  an ad-hoc dependence for the AM diffusion coefficient on rotation core-to-surface ratio. 
 The authors found that a dependence of 
$\Bigl(\Omega_{rad}/\Omega_{env}\Bigr)^3$ reproduces the observations and is consistent 
with what could be expected from an AMRI (Azimuthal Magneto-Rotational Instability instability, R\"udiger et al. 2015).
 Note that this latter dependency- scaling with the differential rotation-  differs from the previous TS one which behaves as 
 $\sim r_{core}^2 \Omega_{core}^2/N_\mu$ near the H-burning shell where the chemical gradient is large.

\paragraph{Conclusion:} 

During the rapid evolution of the PoMS phases, the  AM hydrodynamical (rotation related) transport is 
inefficient to compensate for the strong  core rotation  increase due to the  contraction of the inner layers  with evolution.     
Some missing  AM transport must operate to account for the observations.
While theoretical developments aim at  modelling physically the AM transport candidates, 
other  studies go along another direction and work at  
- characterizing   further the observed rotation profiles namely the location of the internal  differential rotation  and 
- characterizing observationally the missing AM transport as a function of stellar
 parameters and evolutionary phases. These are briefly addressed in the following subsection.

\begin{figure}[t]
 \centering
 \includegraphics[width=9cm,clip]{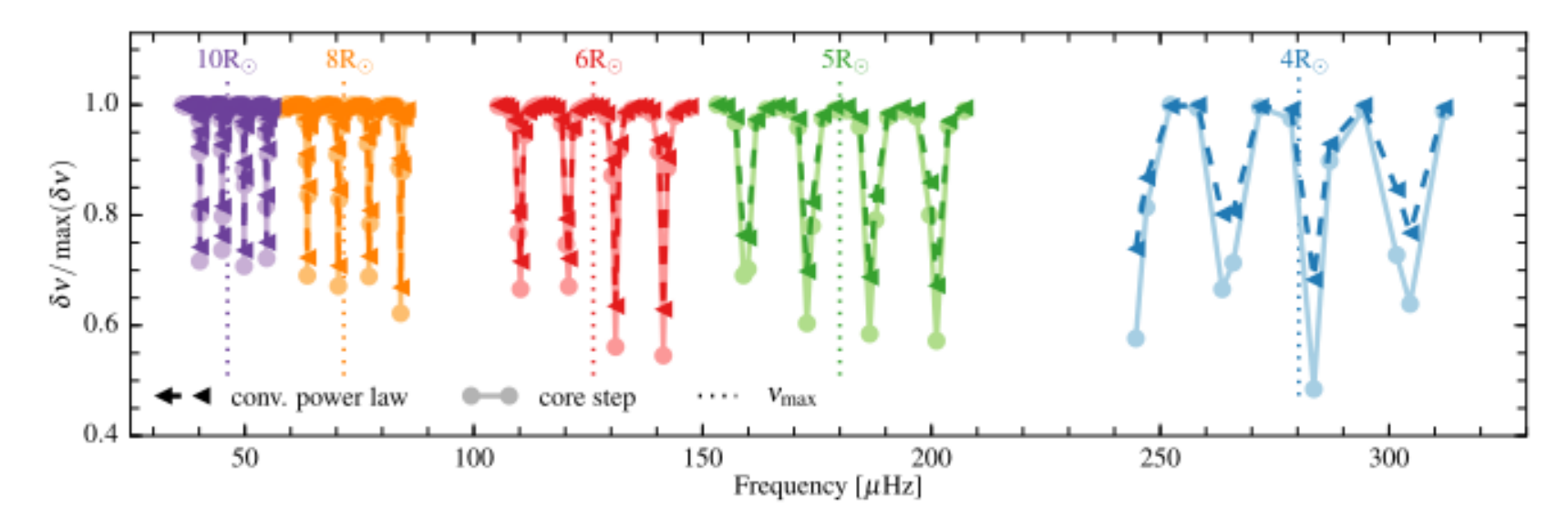}        
 \includegraphics[width=6cm,clip]{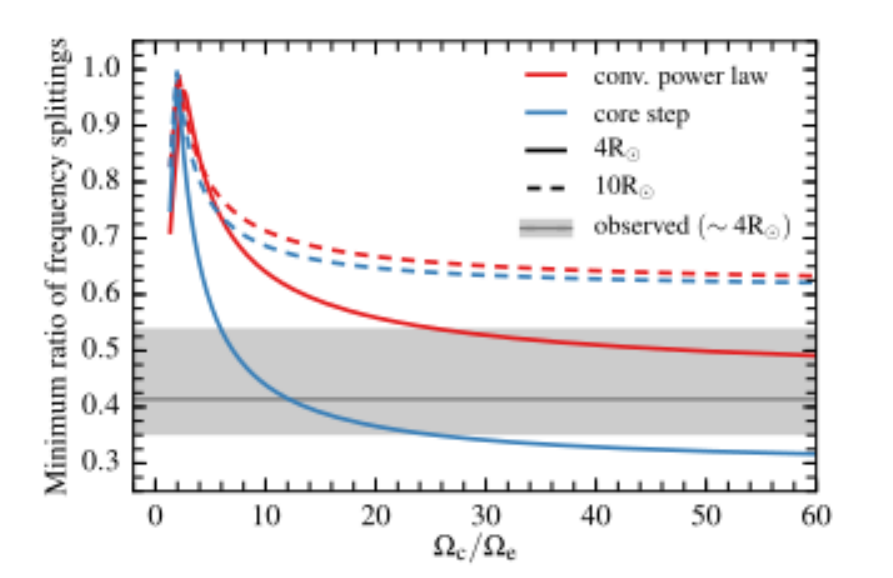}        
   \caption{\label{fig25}  \small \it Left:  rotational  splittings normalized to  their maximal values  
for $1.33 M_\odot$ red giant models  assuming either a two-step rotation profile with the radial  differential rotation just outside of the hydrogen burning 
shell (lighter solid lines) or a rotation profile varying as a power law with radius  in the
convection zone  and uniformly in the radiative region (darker dashed lines). Cases  are labelled with the stellar radius of the model 
for a  proxy of evolution.  
The ratio $\Omega_{env}/\Omega_{core } =7.89$ is taken the same in all cases.  
  The dotted lines show $\nu_{max}$, indicator ot the frequency at maximum power for each
model. Right:   Local splitting minima normalized to the maximum splitting value
 as a function of the rotation ratio $\Omega_{env}/\Omega_{core}$ for several rotation profiles: a two zone model (blue line) and a model assuming a 
uniform rotation in the radiative zone and a power law  decrease in the convective  envelope (red line).
The grey area indicates the range of  observed values.(credit  Klion  \& Quataert , 2017).}
 \end{figure}
 
\subsection{Attempts to characterize further the rotation profile :  location of the radial differential rotation}
 
We have seen that the rotation  core-to-envelope ratio  of PoMs stars is smaller than  expected but remains large. 
    These  stars then maintain some degree of radial differential rotation.  The candidate mechanisms for AM transport in postMS stars
 are able to generate  
 more or less sharp radial rotation gradients which can be located at  different places in the star.   
When  hydrodynamical processes and fluid instabilities dominate, the convective envelope is assumed to  rotate uniformly 
and   a sharp rotation gradient developes near the hydrogen burning shell in the radiative region  as shown in Fig.\ref{fig22}.
 On the other hand, considering magneto-hydrodynamic  instabilities and core-envelope coupling to be responsible for shaping the rotation of red giants, 
 Kissin \& Thompson (2015) studied the case when  the radial differential
rotation resides   in the convection zone  and the radiative interior is rotating uniformly. When  
the impact of the Tayler instability dominates, a  sharp rotation gradient developes  near the hydrogen burning shell in the radiative region 
as for the pure hydrodynamical case but with a much smaller rotation gradient (Fuller et al. 2019). 
 If IGW dominates, the rotation gradient moves outward on secular time scales and  can be catched up  anywhere in the star depending 
at which phase we observe the star (Alvan et al. 2013).
An important hint for finding out  what is  the dominant AM transport mechanism  can then come from the localization 
of the radial differential rotation.  Seismic diagnostics able to locate the rotation gradient  are  therefore of tremendous importance. 
Several methods can be used which indicate - first  that the seismic information is localized   in the central regions and at the very surface; in between not enough information
 with the detected $\ell=1$ modes is available to resolve the rotation gradient in the intermediate regions - second the signature of the location of differential rotation is easier to detect for 
 less evolved red giants. Attention then concentrated on subgiants and early red giants.

The simplest method assumes  a step rotation profile in a two-zone model   
where an average rotation is assumed in the inner part, $\Omega_{core}$, and one average value, $\Omega_{env}$, 
is assumed in  the envelope. In a forward approach, the impact of the position of the rotation gradient 
was then investigated on the rotational splittings  (Deheuvels   2014, 2015, Triana et al 2017).
 Klion \& Quataert (2017)  suggested  a simple seismic diagnostic  to locate the radial differential rotation. 
They showed that at least for the smallest red giants the quantity 
$\delta \nu_{min}/\delta \nu_{max}$ 
evaluated near $\nu_{max}$ is sensitive to whether  the  radial 
differential rotation is  located in the radiative region above the H burning shell  
or concentrated within  the convective envelope.
The splittings of the p-dominated modes $\delta \nu_{min}$ normalized to the maximum splitting $\delta \nu_{max}$ (i.e. of g-dominated modes) 
 indeed are  smallest for  the radial differential rotation located in the radiative region. 
 Fig. \ref{fig25} (left) taken from   Klion \& Quataert (2017)    indicates that the diagnostic  is  more efficient for stars
 with  about 4- 5 stellar radius.
Fig. \ref{fig25}   (right)  plots $\delta \nu_{min}/\delta \nu_{max}$ as a function of 
$ \Omega_{core}/\Omega_{env}$ and  
 illustrates the fact that for most early giant stars   the assumption of the  radial 
differential rotation located in the radiative region is favoured.

Using Eq.\ref{dnumindnumax} with various rotation profiles, it is possible to explain the decrease
 of $\delta \nu_{min}/\delta \nu_{max}$ with  the rotation 
core-to-envelope (Fig. \ref{fig25} (right)) and the signature of a different location of the differential rotation. As in Klion et al. (2017), I consider two rotation profiles. 

\begin{itemize}
\item A first rotation profile assumes that the rotation gradient is located within the g-cavity at a radius $r_i$ ($ r_a<r_i<r_b$) and is  uniform below and above
 that-is 
$$ \Omega(r) = \{ 
\begin{matrix}
 & \Omega_c ~ & {\rm for} ~ r \leq r_i < r_b  \\
 & \Omega_e ~ & {\rm for} ~  r_i \leq r  
\end{matrix} 
$$

The core rotation averaged over the g -cavity and the envelope rotation averaged over the p-cavity can be respectively approximated  as (Goupil et al, 2013)
\begin{eqnarray}
\Omega_{core} &=&  \frac{1}{ \int_{r_a}^{r_b}~ \frac{N}{r} dr }  \int_{r_a}^{r_b}~ \Omega \frac{N}{r} dr \\
\Omega_{env} &=&  \frac{1}{ \int_{r_c}^{r_d}~ \frac{dr}{c_s}   }  \int_{r_c}^{r_d}~ \Omega \frac{dr}{c_s}  
\end{eqnarray}

For the present purpose, it is enough to assume that the Br\"unt -V\"aiss\"al\"a frequency in the g-cavity is $N\sim N_{core}=$ const 
and that in the p-cavity  the inverse of the sound speed dominates 
close to the surface $1/c_s \sim 1/c_{R}=$const.  One then obtains 
\begin{eqnarray}
\Omega_{core} &\sim &  \Omega_c ~\frac{\log(r_i/r_a)}{\log(r_b/r_a)} \\
\Omega_{env} &\sim& \Omega_e    \\
\end{eqnarray}
This gives for the rotation ratio
$$ {\cal R}=\frac{\Omega_{env}}{\Omega_{core}}  = \frac{\Omega_e}{\Omega_c}  ~\frac{\log(r_b/r_a)}{\log(r_i/r_a)} $$
and for the splitting ratio (Eq.\ref{dnumindnumax}):
$$ \frac{\delta \nu_{min}}{\delta \nu_{max}}
\sim  \frac{1}{1+1/(q{\cal N})}  ~ \Bigl(1+ \frac{2}{q{\cal N}}   \frac{\Omega_{env}}{\Omega_{core}}  \Bigr) 
= \frac{1}{1+1/(q{\cal N})}  ~     \Bigl(1+ \frac{2}{q{\cal N}}   \frac{\log(r_b/r_a)}{\log(r_i/r_a)} ~\frac{\Omega_{e}}{\Omega_{c}} \Bigr)  $$
The splitting ratio linearly increases with 
$ \Omega_{e}/\Omega_{c}$  or when considering the  $\Omega_c/\Omega_e$ ratio  as in Klion et al. (2017)
(Fig.\ref{fig25})  decreases with $\Omega_{c}/\Omega_e $  as  $1/(\Omega_{c}/\Omega_e)$ with a rate 
$$ rate_1 =\frac{2}{q{\cal N}}   \frac{\log(r_b/r_a)}{\log(r_i/r_a)} \sim \frac{2}{q{\cal N}}$$
The last equality is obtained for the 4$R_\odot$ model of Klion et al. (2017) for which  one has for the radii normalized to the stellar radii 
$ r_a << 10^{-4}; r_i \sim 0.01 ; r_b \sim 0.1; =r_{BZC}\sim 0.2;  r_c \sim 0.3, r_d \sim 1 $.
Since $r_a << r_i<<r_b$, so that $\ln(r_b/r_a)/\ln(r_i/r_a)>1$ goes to 1 when $r_a$ goes to zero.

\item For a rotation gradient in the convective region, a simplified rotation profile can be assumed of the form
$$ \Omega(r) = \{ 
\begin{matrix}
 & \Omega_c ~ & {\rm for} ~ r \leq   r_{BZC}  \\
 & \frac{\Omega_e}{r^\alpha}  ~ & {\rm for} ~  r_{BZC}  < r  
\end{matrix} 
$$
For sake of simplicity, $\alpha=1$ hereafter, then 
\begin{eqnarray}
\Omega_{core} &\sim &  \Omega_c  \\
\Omega_{env} &\sim& \Omega_e ~ \frac{\ln(1/r_{BZC})}{1-r_{BZC}}   \\
\end{eqnarray}
This gives for the rotation ratio
$$ {\cal R}=\frac{\Omega_{env}}{\Omega_{core}}  = \frac{\Omega_e}{\Omega_c}  ~\frac{\ln(1/r_{BZC})}{1-r_{BZC}}  $$
and for the splitting ratio (Eq.\ref{dnumindnumax}):
$$ \frac{\delta \nu_{min}}{\delta \nu_{max}}
\sim     \frac{1}{1+1/(q{\cal N})}  ~     \Bigl(1+ \frac{2}{q{\cal N}}   ~\frac{\ln(1/r_{BZC})}{1-r_{BZC}}    ~\frac{\Omega_{e}}{\Omega_{c}} \Bigr)  $$
The splitting ratio again linearly increases with 
$ \Omega_{e}/\Omega_{c}$  or when considering the  $\Omega_c/\Omega_e$ ratio  as in Klion et al. (2017)
  decreases with $\Omega_{c}/\Omega_e $  as  $1/(\Omega_{c}/\Omega_e)$. 
 with a rate 
$$ rate_2= \frac{2}{q{\cal N}}  ~\frac{\ln(1/r_{BZC})}{1-r_{BZC}}\sim \frac{2}{q{\cal N}}  ~2$$
As above, the last equality is obtained for the 4$R_\odot$ model of Klion et al. (2017).
\end{itemize}
The ratio $rate_2/rate_1 \sim  2$  indicates that the splitting ratio $\delta \nu_{min}/\delta \nu_{max}$ 
decreases  from 1 with $\Omega_{c}/\Omega_e$   more rapidly for the core step rotation profile than for a   convective
power law dependence of the rotation. We recover the results found numerically by Klion et al. (2017). 

Note that a differential rotation causes $\delta \nu_{max}\approx \Omega_{core}/(2\pi)$ to take also different values for a same 
$\Omega_e/\Omega_c$  ratio. This means that a least in theory a plot $\delta \nu_{min}/\delta \nu_{max}$ as a function of $\delta \nu_{max}$
ought to discriminate between two different differential rotations. This will be tested with a large sample of stars with similar masses and radii 
 to try to reach some general  conclusions  about the location of the rotation gradient. 
 An independent measure of the surface rotation such as that obtained with a rotational modulation of the light curve would of course be invaluable 
in that respect. 

\medskip
 More sophisticated methods such as inversion of the rotational splittings
  can also be used (Deheuvels et al. 2012, 2014, 2015; Di Mauro et al.,  2016; Triana et al., 2017). 
For instance,  inversions performed by Deheuvels et al. (2014) showed that  
a sharp variation of the rotation profile near the H burning shell 
is favored over a smooth profile at least for two Kepler early red giants.  
   Triana et al. (2017)   used the  rotational splittings of 13 out of the 19 red giants analyzed by Corsaro et al. (2015)
 to derive their  average rotation rates. The stars are in a similar evolutionary stage (with a  radius $R\sim 5 R_\odot$)  and
 their core - to - surface rotation ratios are found in the range 5-10.
    For one star,  Triana et al. (2017)   derived the rotation profile using  the various aforementionned  methods 
and found that they essentially give the same results and concluded that -with  observations currently at
at hand-  no precise information can be obtained in the  radiative region above the H-burning shell. The authors also conclude that 
the Klion \& Quataert (2017) diagnostic is inefficient to distinguish between a radial differential rotation either in 
the radiative or in  the convective region for the considered stars.  
 
\begin{figure}[t]
 \centering
 \includegraphics[width=14cm ]{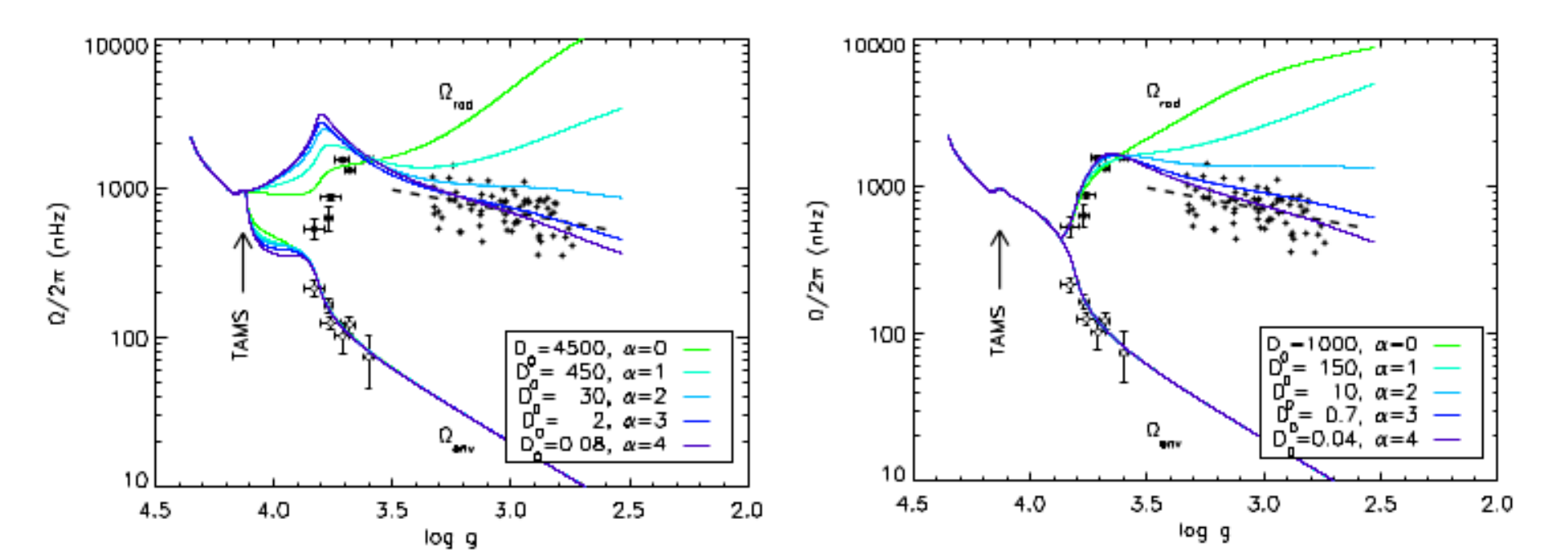}        
   \caption{\label{fig29}  \small \it  right:
  evolution of the core rotation rate assuming  a AM diffusive transport coefficient scaling as  $\Bigl(\Omega_{rad}/\Omega_{core}\Bigr)^\alpha$
and  a enforced uniform rotation   during the main sequence (left) and maintained on the subgiant phase  until slightly before 
the first dredge starts (credit Spada et al. (2016)). }
 \end{figure}

\subsection{Attempts to characterize the underlying AM transport process : efficiency of
the missing mechanism(s)}

A series of work  was carried out  with the aim of characterizing the efficiency of the unidentified AM transport mechanism operating in PoMS
 stellar interiors.  This was achieved by including an additional constant  viscosity $\nu_{add}$ to 
   the AM transport equation 
i.e. $\nu_v = \nu_{shear} + \nu_{add}$ in  Eq.\ref{flux1} and   ajusting the values  $\nu_{add}$ in order to  reproduce the observations. 
Such studies were performed for 
 a $1.5 M_\odot$ red giant KIC 8366239   (Eggenberger et al. 2012),  a $0.84 M_\odot$  red giant
KIC 7341231 (Eggenberger et al.,  2015)  as well as for the observed  subgiants (Eggenberger et al., 2018) . 
It was found that when constraints on both the core and surface rotation rates  are available- which is particularly true for subgiants- 
a precise value can be obtained for $\nu_{add}$. Indeed  
while the knowledge of the core rotation rate imposes a sufficiently high value of $\nu_{add}$ so that it dominates the AM transport, 
a constraint on  the surface rotation  rate  limits the maximal value of this added viscosity.  
 These  precise determinations led to the conclusions that 
 the efficiency of the missing AM transport as represented by $\nu_{add}$ must increase with the mass in order to reproduce the observations.
 For instance, $\nu_{add}$ for KIC 7341231 is at least three times lower than the one obtained for KIC 8366239
independently of their different evolution stage. 
Using the observations of a large sample of red giant stars  (Mosser et al.,  2012) and a sample of subgiant stars (Deheuvels, 2015), 
 it was also found that  the behavior of $\nu_{add}$ with evolution differs  during the subgiant and the red giant branch: 
 the efficiency decreases on the subgiant branch and increases during 
the red giant ascending branch (Cantiello et al., 2014; Eggenberger et al.,  2015; Spada et al., 2016, Eggenberger et al., 2018).
 This conclusion   closely follows what is expected  from the observations when the core rotation 
of the subgiants increases whereas the core of the ascending red giant decreases despite the core contraction (Deheuvels 2015). 
  Another important information brought by these studies is  that  the value of $\nu_{add}$ is not sensitive to the past
 rotation history of the star, except
 at the very beginning of the subgiant phase  and can therefore probe the AM transport taking place  during the post MS phase.

In another approach, Spada et al. (2016) computed  the evolution of the rotation profile of 
a 1.25 stellar model  up to the RGB  phase including internal AM transport as a pure diffusive transport ($U_2=0$) in Eq.\ref{flux1}. 
Surface AM losses were taken into account  according to the Kawaler (1988)'s prescription.  
On one hand, the Kawaler's constant $K_w$ is adjusted so that the models can reproduce the observed envelope rotation of the subgiants and red giants. 
On the other hand, the authors assumed a turbulent diffusion for the internal AM transport of the form  
$$D=D_0 \Bigl(\frac{\Omega_{rad}}{\Omega_{core}}\Bigr)^\alpha $$ 
where $D=\nu_v$ in our notation (Eq.\ref{flux1}).
The exponant $\alpha$  was determined so that the evolutionary stellar models can reproduce  
the observed core rotations of red giants and their evolution.   
 Motivated by the fact that  observations of main sequence stars indicate a rigid rotation
 for several stars besides the Sun (Nielsen et al., 2014; Benomar et al., 2015),   
Spada et al (2016) studied first the case when   a rigid rotation is enforced on the main sequence until the TAMS. 
The best match for the red giant core evolution is obtained for a   value 
$\alpha=3$, which as already mentioned (Sect.4.2) may have some physical justification. 
However, the evolution of the core rotation on  the subgiant phase is not reproduced correctly: 
the core accelerates too rapidly (Fig.\ref{fig29}(left)).  
In a second computation, the rigid rotation is imposed until a point where the core rotation can be well reproduced for all subgiants. 
Spada et al. (2016) 
identified this time as roughly the time when the hydrogen burning shell has been fully extended before the convective envelope start to recede. 
Then a single set of parameter values for  ($\alpha, D$) is able to reproduce the evolution of the subgiant and red giant branches (Fig.\ref{fig29}(right)).
 The data for the red giants  were  taken from Mosser et al. (2012),  
some small ajustement of the parameter values might be necessary in order to agree  with the smaller spin down found by Gehan et al. (2018).

\begin{figure}[t]
 \centering
 \includegraphics[width=8cm,clip]{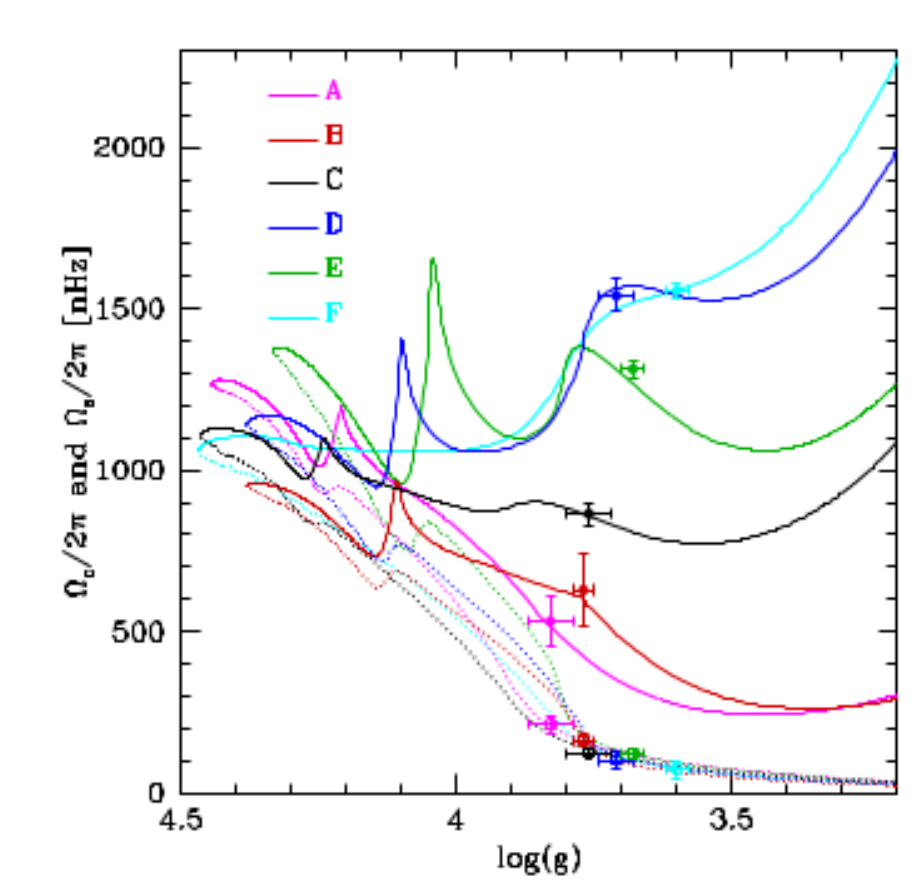}        
 \includegraphics[width=8cm,clip]{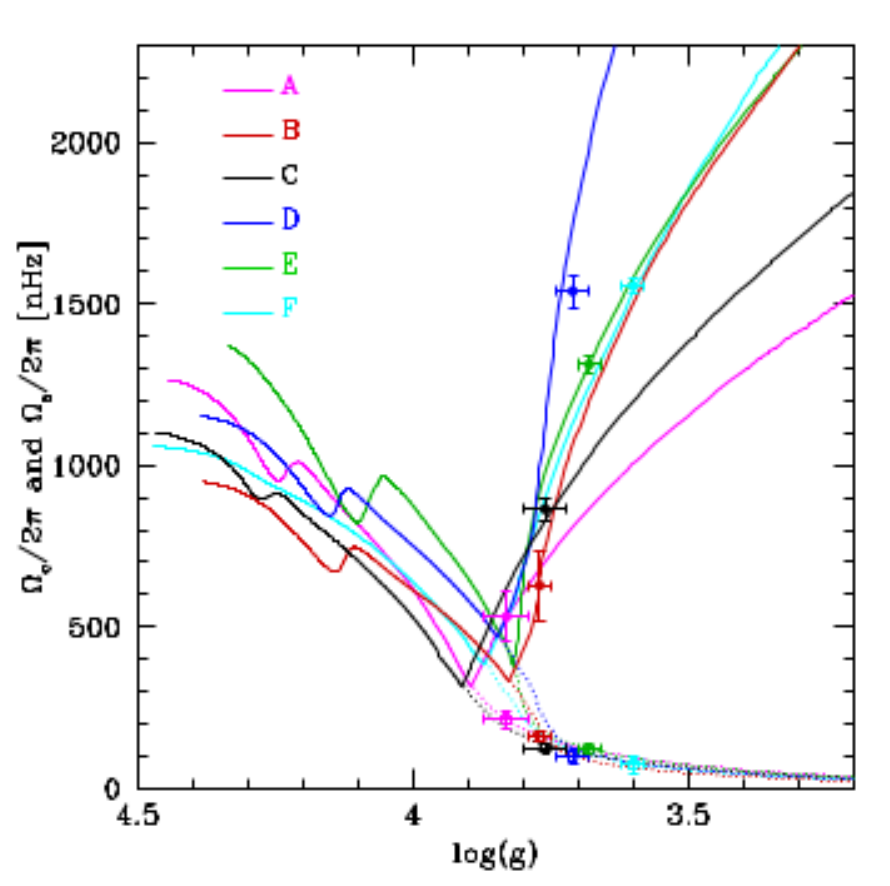}        
   \caption{\label{fig50}  \small \it  left:  Core (solid lines) and surface (dotted lines) rotation rates
as a function of gravity for rotating models of the six subgiants determined by Deheuvels et al. 2014.
Dots and open circles indicate the seismic  values of
core and surface rotation rates determined for the six subgiants. 
Right same as left but the models are computed by assuming solid-body rotation
until a given point during the post-main sequence. (Credit Eggenberger et al., 2019) }
 \end{figure}

Eggenberger et al. (2018) find a similar behavior when computing the AM  transport according to Eq. \ref{AM} including meridional circulation
 and shear induced turbulence
and including an  additional viscosity $\nu_{add}$ as mentionned previously. 
The values of the additional viscosity  and the rotation period at the ZAMS are ajusted  for each subgiant so that the core and 
surface rotation  of the models   reproduce simultaneously the observed ones. The additional viscosity  is added  first from 
the very of beginning of the evolution of the model (Fig.\ref{fig50}(left)). 
Following the suggestion  by Spada et al. (2016), Eggenberger et al. (2018) also  evolved the  models for each subgiant 
 enforcing a rigid rotation on the MS and  beyond  the TAMS until a point where the rigid rotation is turned off 
and the contracting core is free to accelerate up to the observed value. 
In that case, no additional AM transport than meridional circulation and shear turbulence 
is necessary  on the subgiant branch (Fig.\ref{fig50}(left)). The issue here is that according to the authors, 
 the  time at which the rigid rotation needs to be turned off differs for each subgiant and  does seem to appear at a specific phase of the subgiant evolution.

\section{Conclusions}
 
Significant advances in Galactic archeology were brought up by the ability of the seismology of red giants  to  constrain precisely 
 their masses and ages. At the same time, 
 the seismic probing of the internal properties  of these stars  revealed   severe flaws in our understanding and modelling of 
the structure and the internal AM evolution of the  red giants. This motivated a vivid brainstorming activity in the community  
leading to  several valuable advances in the field.  In this  review,
I chose to address two important issues:  mixing beyond the convective core of red clump stars and angular momentum transport in low mass post main sequence stars. 
In a nutshell, what can be concluded?

On the structural side, the observed period spacings  of red clump stars  confirm that these stars have quite 
larger, more massive convective cores than predicted by the classical Schwarzschild criterion as implemented in many stellar evolution.
The properties of the extra-mixing region above the formal convective core  can in principal be determined as well. The contribution of the 
Li\`ege school led by Arlette Noels has been  determinant in the significant advances of the past decade  for that challenging field. 
However the way forward from the seismic observations to a proper modelling of the central regions of CHeb stars
 is not easy and the final word is not said yet.   To proceed beyond a mere ajustement of a free parameter, one must await 
significant advances from numerical simulations and
 theoretical developments.  On the observational side, the previous studies have shown 
the necessity of a larger sample of stars to cover  the range of mass, evolution and metallicity.

The general picture of AM transport is nowadays certainly more satisfying since 
  it is now possible to reconciliate the core rotation rates  of the red giant stars  and the rotation rates of  their descendants, the white dwarfs.
Several processes  have been identified as likely  candidates  for the AM evolution of PoMS star. 
However a tremendous amount of work  is still awaiting ahead of us  to confirm these ideas, this must be a combination of
theoretical developments, numerical simulations and data analyses.  
 Indeed a full modelling of the total AM flux along the evolution must now include all these processes which might operate and dominate 
at different phases of the star evolution 
$ {\cal F}_{tot} = {\cal F}_{circ}+  {\cal F}_{shear}+  {\cal F}_{waves}+{\cal F}_{modes} +{\cal F}_{B}$.
  Prescriptions for the transport coefficients for each individual process are nowadays included and studied individually on more 
physical grounds but  certainly deserve further  improvements. The interations between these processes ought now to be taken into account 
and quantified  by their consistent implementation in 1D stellar models.  
Morover, whereas most of the identified  AM transport  processes might not have a  significant direct impact on the structure and on the frequencies of 
the axisymmetric $m=0$ modes, the cumulative effect of each dominant process during the evolution of the PoMS might result in some chemical,
 hence structural, modifications which remain to be investigated and their seismic signatures determined. In that sense,   only 
 a few  individual stars have been studied so far.  On the side of 
 seismic theoretical developements, promising  prospects are  the study of glitches  (Cunha et al.,  2015),  
of the information that $\ell=2$ modes can carry out (Deheuvel et al., 2017), the case of evolved fast rotators (Ouazzani et al.,  2013) 
and the  series of investigation on the coupling factor  which  characterizes the evanescent region between the g and p resonant cavities 
(Pincon et al., 2019 and reference therein).

While the data from Kepler/K2  have certainly still much more  to tell,
 the currently running NASA mission, TESS (Schofield et al., 2019)  and the ESA project PLATO  (launch in 2026, Rauer et al., 2014) 
will offer  unvaluable opportunities to  address further  the above issues.

\section*{Acknowledgements}
It was a pleasure to attend the Liege conference for the 75th birthday of Arlette Noels. 
The  present contribution  is then dedicated to my dear friend and highly esteemed colleague Arlette. I truely thank Arlette for  our lively scientific discussions,
 her  deep understanding of the physics and evolution of stellar interiors  and her pedagogical talent to transmit  her knowledge       
and most importantly for our good time  and laughs around a diner table.  I hope she will forgive me for the involuntary imperfections 
of the present paper.  
I also gratefully thank Marc -Antoine Dupret for his careful reading of the manuscript.
\footnotesize
\beginrefer

\refer  Aizenman, M., Smeyers, P., Weigert, A., 1977, AA 58, 41 
  
\refer  Alvan, L.,  Mathis, S., Decressin, T., 2013b,  AA 553, A8
  
\refer Amard, L., Palacios, A., Charbonnel, C. and 2 co-authors, 2016, AA 587, 105
    
\refer Andr{\'e}, Q., Mathis, S., Amard, L., 2018, sf2a. conf 145

\refer Arnett, W.~D., Meakin, C., Hirschi, R. et al. 2018, arXiv181004653
 
\refer Arnett, W.~D., Meakin, C., Viallet, M. et al., 2015, ApJ., 809, 30

\refer Baglin, A. and the CoRoT team, 2016, 
The CoRoT Legacy Book: The adventure of the ultra high precision photometry from space, by the CoRot Team, 
  ISBN: 978-2-7598-1876-1, EDP Sciences,  p.1 

\refer  Baglin, A., Auvergne , M. , Barge, P. et al., 2006, ESASP 1306, 33

\refer Beck, P. G., Montalban, J., Kallinger, T.   et al., 2012, Natur, 481, 55

\refer Bedding, T. R., Mosser, B., Huber, D. et al, 2011, Natur, 471, 608 

 \refer   Belkacem, K.,  Marques, J.P.,  Goupil, M.J.,  and 6 co-authors, 2015a, AA 579, A30

 \refer   Belkacem, K., Marques, J. P., Goupil,  M.J., and 6 co-authors, 2015b, AA 579, A31

\refer  Benomar, O., 2013  in Progress in Physics of the Sun and Stars: A New Era in Helio- and Asteroseismology
ASP Conference Series 479, 161 H. Shibahashi and A. E. Lynas-Gray, eds.

\refer  Benomar, O.,  Belkacem, K.,   Bedding T. R. et al., 2014, 781, L29 

\refer Benomar, O., Takata, M.,, Shibahashi et al., 2015, MNRAS 452, 2654

\refer Borucki , William J., 2018, 
 in Handbook of Exoplanets, ISBN 978-3-319-55332-0. Springer International Publishing AG, part of Springer Nature 

\refer Borucki , W. J., Koch, D., Basri, G. et al., 2010, Sci. 327, 977

\refer Bossini, D., Miglio, A.,  Salaris, M.  et al.,  2017, MNRAS, 469, 4718

\refer  Bossini, D., Miglio, A., Salaris, M. et al., 2015, MNRAS 453, 2290

\refer Brun,, A.~S., Palacios, A., 2009, ApJ 702, 1078

\refer  Buysschaert, B., Beck, P.~G., Corsaro, E. et al.,2016, A\&A 588, 82

\refer Cantiello , M., Mankovich, C., Bildsten , L. et al., 2014, ApJ 788, 93

\refer Ceillier, T., Eggenberger, P., Garc{\'\i}a, R.~A. , Mathis, S., 2013 AA 555, 54

\refer  Ceillier, T., Tayar, J., Mathur, S. et al., 2017, AA 605, 111
       
\refer  Chaplin, W. J., Miglio, A., 2013, ARA\&A 51, 353 

\refer  Charpinet, S., Giammichele, N., Zong, W. et al., 2018, OAst 27, 112

\refer Christensen-Dalsgaard, 2007, Lecture notes on Stellar Oscillations

 \refer  Corsaro, E., De Ridder, J., Garc{\'\i}a, R.~A.,2015, AA 579, 83

\refer  Constantino, T., Campbell, S. W., Christensen-Dalsgaard, J. et al., 2015, MNRAS 452, 123

 \refer Constantino, T., Campbell, S. W., Lattanzio, J. C.,2017, MNRAS 472, 4900

\refer   Cox, J. P., 1980, 'Theory of stellar pulsation',  

\refer Cunha, M.~S., Stello, D., Avelino, P.~P. et al., 2015, ApJ 805, 127
 
\refer Deheuvels, S. Garcı́a, R. A.  Chaplin, W. J., 2012, ApJ 756, 19 

\refer Deheuvels, S. Dogan, G. Goupil, M. J., 2014,   AA 564, A27 

 \refer Deheuvels, S., Ballot, J.,  Beck, P. G.,  2015, AA 580, A96

\refer Deheuvels, S., Ouazzani, R.~M., Basu, S.,2017, AA 605, 75

\refer Di Mauro, M. P., Ventura, R., Corsaro, E. et al., 2018, ApJ 862, 9
  
\refer Di Mauro, M.~P., Ventura, R., Cardini, D. et al., 2016, ApJ 817, 65

\refer Dupret M.-A., Belkacem K., Samadi R. et al., 2009, AA 506, 57

\refer  Dziembowski, W.~A. , 1971, AcA 21, 289

\refer Dziembowski, W.~A.; Gough, D.~O.; Houdek, G.  et al., 2001, MNRAS 328, 601

\refer Eggenberger, P., Deheuvels, S., Miglio, A. et al., 2019, A\&A 621, 66,

\refer Eggenberger, P., Lagarde, N., Miglio, A. et al.,  2017, A\&A 599, 18
      
\refer Eggenberger , P., 2015, EAS 73, 26
  
\refer Eggenberger, P., Montalb{\'a}n, J., Miglio, A., 2012, A\&A 544, 4

 \refer   Endal, A.~S. \& Sofia, S., 1978, ApJ 220, 279

 \refer   Endal, A.~S. \& Sofia, S., 1981, ApJ 243, 625

\refer  Fuller, J., Lecoanet, D., Cantiello, M. et al., 2014, ApJ 796, 17
 
\refer  Fuller, J., Piro, A. L., Jermyn, A. S., 2019 MNRAS 485, 3661

\refer Gabriel, M., Noels, A., Montalb{\'a}n, J. et al.,  2014, A\&A 569, 63

\refer  Gehan, C., Mosser, B., Michel, E. et al., 2018, AA 616, 24

\refer  Goldstein, J., Townsend, R.~H.~D., Zweibel, E.~G., 2018, arXiv180808958

 \refer  Gough, D.~O.,1993, 
in 'Astrophysical Fluid Dynamics - Les Houches 1987', elsevier science publisher, p. 399
  
\refer Goupil, M.~J., Mosser, B., Marques, J.~P. et al.,  2013, AA 549, 75
 
\refer Kawaler, S. D.,    1988, ApJ 333, 236

\refer Hekker, S., Christensen-Dalsgaard, J.,2017 ,AAR 25, 1

\refer Hekker, S., Elsworth, Y., Angelou, G.~C. et al., 2018, AA 610, 80
       
\refer Hermes, J.~J., G{\"a}nsicke, B.~T., Kawaler, S. D. et al.,  2017, ApJS 232, 23

\refer Jiang, C. \& Christensen-Dalsgaard, J., 2014, MNRAS 444, 3622

\refer Jouve, L., Gastine, T., Ligni{\`e}res, F., 2015, AA 575, 106
 
\refer Kissin  Y.,  Thompson  C., 2015, ApJ, 808, 35
    
\refer Klion, H. \& Quataert, E., 2017, MNRAS 464, 16

\refer  Lecoanet, D. \& Quataert, E., 2013, MNRAS 430 2363
     
\refer Loi, S. T. \& Papaloizou, J. C.~B., 2018, MNRAS 477 5338

\refer Maeder, A., 2009,  
 'Physics, Formation and Evolution of Rotating Stars', Springer

\refer Maeder, A. \& Meynet, G., 2004, AA 422, 225

 \refer Maeder, A. \& Meynet, G., 2005, AA 440, 1041
 
\refer Maeder, A. \&  Meynet, G., 2014, ApJ 793, 123
 
\refer Maeder, A., Meynet, G., Lagarde, N. et al., 2013, AA 553, 1

\refer   Maeder, A. \& Zahn, J.P., 1998, AA 334, 1000
 
\refer   Marques, J.~P., Goupil, M.~J., Lebreton, Y. et al., 2013, AA 549, 74

\refer Mathis, S., Zahn, J.-P., 2004, AA 425, 229

\refer Mathis, S, 2013, Transport Processes in Stellar Interiors, LNP 865, 23

\refer Mathis, S \& Alfvan,  2013, Transport Processes in Stellar Interiors, EAS 63, 269

\refer  Mathis, S., Prat, V., Amard, L. et al., 2018, A\&A 620, 22
  
\refer  Meynet, G., Ekstrom, S., Maeder, A. et al.,  2013, LNP 865, 3

\refer  Montalb{\'a}n , J., Miglio, A., Noels, A. et al., 2013, ApJ 766, 118

\refer Montalb{\'a}n, J., Noels, A., 2014,  in 'Asteroseismology of  red giant stars: The potential of dipole modes ',
 EPJWC 43   

\refer Mosser, B., Barban, C., Montalb\`an, J., et al. 2011, AA, 532, A86 
 
\refer Mosser,  B.,  Goupil, M. J., Belkacem, K., 2012a, AA 540, A143
 
\refer Mosser, B.,   Goupil,  M. J.  Belkacem, K., 2012b, AA 548, A10

\refer  Mosser, B.,, O. Benomar, K. Belkacem, 2014, AA 572  

\refer   Mosser, B., Vrard, M., Belkacem, K. et al., 2015, AA 584, 50
     
\refer  Mosser, B., Miglio, A. and CoRot Team, 2016,
in The CoRoT Legacy Book: The adventure of the ultra high precision photometry from space, by the CoRot Team .
 ISBN: 978-2-7598-1876-1.    EDP Sciences , p.197
        
\refer  Mosser, B., Pin{\c{c}}on, C., Belkacem, K. et al., 2017, AA 607, 2

\refer Mosser, B., Gehan, C., Belkacem, K. et al., 2018, AA 618, 109

\refer Nielsen, M.~B., Gizon, L., Schunker, H. et al.,  2014, A\&A 568, 12

\refer Noels, A., Montalban, J., Miglio, A. et a., 2010, ApSS 328, 227

\refer Noels, A., 2013, LNP 865, 209

\refer  Ouazzani, R. -M., Goupil, M. -J., Dupret, M. -A. et al., 2013, ASPC 479,  335

\refer Palacios, A., 2013, EAS 62, 227

\refer Paxton, B.,  Schwab, J., Bauer, E. B. et al., 2018, ApJS 234, 34

\refer Paxton, B., Smolec, R., Schwab, J.  et al., 2019, arXiv190301426P,
   
\refer  Pin{\c{c}}on, C., Belkacem, K., Goupil, M.~J., 2016, AA 588, 122

\refer    Pin{\c{c}}on, C., Belkacem, K., Goupil, M.~J. et al., 2017, AA 605, 31
         
\refer   Pin{\c{c}}on, C.,Takata, M., Mosser, B., 2019, arXiv190505691

\refer Pinsonneault, M.~H., Kawaler, S. D., Sofia, S. et al., 1989, ApJ 338, 424

\refer  Press, W.~H., 1981, ApJ 245, 286

\refer  Rauer, H., Catala, C., Aerts, C. et al., 2014, ExA 38, 249

\refer R{\"u}diger, G., Gellert, M.,Spada, F. et al., 2015, AA 573, 80

\refer  R\"udiger, G., Schultz, M., Kitchatinov, L.~L.,2014, arXiv1404.3562

\refer  Salaris, M. \& Cassisi, S., 2017, RSOS 4, 170192 
 
\refer Schatzman, E., 1993, AA 279, 431

\refer Scuflaire, R. ,1974, AA 34, 449,
 
\refer Schofield, Mathew, Chaplin, W. J., Huber, D. et al., 2019, ApJS 241, 12

\refer Shibahashi, H., 1979, PASJ 31, 87

\refer  Spada, F., Gellert, M., Arlt, R. et al., 2016, AA 589, 23

\refer  Spruit, H.~C., 1999, AA 349, 189
 
\refer Spruit, H.~C.,2002, AA 381, 923

\refer  Spruit, H.~C.,2015, AA 582, L2

\refer Takata, M.,  2016a, Publ. Astron. Soc. Japan   68, 6

\refer Takata, M., 2016b, Publ. Astron. Soc. Japan 68, 91

\refer Talon, S. \& Zahn, J. -P., 1997, AA 317, 749

\refer Talon, S., 2004, IAUS 215, 336

\refer Talon, S., 2008, EAS 32, 81  

\refer Talon, S., \& Charbonnel, C. 2005, AA, 440, 981

\refer Talon, S. \& Charbonnel, C., 2008, AA 482, 597
   
\refer  Tayar, J., Ceillier, T., García-Hernández, D.A., 2015, ApJ 807, 82  

\refer Townsend, R., 2014, in IAU Symp. 301, eds. J. A. Guzik, W. J. Chaplin,
G. Handler,  A. Pigulski, 153

\refer Triana  S. A., Corsaro  E., De Ridder J. et al., 2017, AA, 602, A62

\refer   Unno, W., Osaki, Y., Ando, H., Saio, H., Shibahashi, H. , 1989
 Nonradial oscillations of stars, second edition , University of Tokyo press.

\refer    Vrard, M., Mosser, B., Samadi, R. et al., 2016, AA 588, 87


\refer Zahn, J.P., 1991, AA,   252, 179
    
\refer Zahn, J. P., 1992, AA 265 115

\refer  Zahn, J.P., Talon, S., Matias, J., 1997, AA 322, 320

\refer  Zahn, J.P., 2013, EAS 63, 245 
     
\endrefer

\end{document}